\documentclass[letter,11pt]{article}
\pdfoutput=1

\usepackage{setspace}
\usepackage{jheppub}
\usepackage{amsmath,amssymb,array,calc,rotating,epsfig,psfrag, amscd, datetime, microtype,mathtools}
\usepackage{verbatim}
\usepackage{bm}
\usepackage{upgreek} 

\usepackage[title,titletoc,toc]{appendix}
\usepackage{tikz-cd}    							

\usepackage{titlesec}
\titleformat{\paragraph}[runin]{\normalfont\normalsize\itshape}{\theparagraph}{}{}[.] 
\titlespacing{\paragraph}{0pt}{0pt}{*1}

\makeatletter
\gdef\@fpheader{\ }                    
\makeatother

\makeatletter
\g@addto@macro\bfseries{\boldmath}
\makeatother

\newcommand\qqq{\qquad\qquad}

\newcommand{\quotient}{/}							
\newcommand{\qquotient}{/\!\!/}				
\newcommand{\qqquotient}{/\!\!/\!\!/}	
\DeclareMathOperator{\End}{\text{End}}
\newcommand{\ext}{\mbox{\large $\wedge$}} 
\newcommand{\eqspace}{\mathrel{\phantom{=}}{}} 

\newcommand{\R}{\mathbb{R}}

\newcommand{\ee}{\text{e}}
\DeclareMathOperator{\vol}{\text{vol}}

\newcommand{\dd}{\text{d}}

\newcommand{\ii}{\text{i}}
\newcommand{\del}{\partial}

\newcommand{\ft}{\tilde{f}}
\newlength{\sswidth}

\DeclareMathOperator{\ad}{\text{ad}}
\DeclareMathOperator{\im}{\text{Im}}
\DeclareMathOperator{\re}{\text{Re}}

\newcommand{\GL}[1]{\text{GL}(#1)}
\newcommand{\SL}[1]{\text{SL}(#1)}
\newcommand{\SU}[1]{\text{SU}(#1)}
\newcommand{\Uni}[1]{\text{U}(#1)}
\newcommand{\SUstar}[1]{\text{SU}^{*}(#1)}

\newcommand{\USp}[1]{\text{USp}(#1)}
\newcommand{\Orth}[1]{\text{O}(#1)}
\newcommand{\SO}[1]{\text{SO}(#1)}
\newcommand{\Ex}[1]{\text{E}_{#1}}
\newcommand{\Fx}[1]{\text{F}_{#1}}
\newcommand{\Gx}[1]{\text{G}_{#1}}

\newcommand{\su}[1]{\mathfrak{su}_{#1}}

\newcommand{\ex}[1]{\mathfrak{e}_{#1}}

\newcommand{\gl}[1]{\mathfrak{gl}_{#1}}
\newcommand{\sln}[1]{\mathfrak{sl}_{#1}}

\newcommand{\Dorf}{\hat{L}}
\newcommand{\Dorft}{L}

\newcommand{\rep}[1]{\boldsymbol{#1}}

\newcommand{\id}{\mathbb{I}}

\newcommand{\lrc}{\lrcorner}

\DeclareMathOperator{\tr}{\text{tr}}

\newcommand{\bbR}{\mathbb{R}}
\newcommand{\bbC}{\mathbb{C}}

\newcommand{\cA}{\mathcal{A}}

\newcommand{\cL}{\mathcal{L}}
\newcommand{\cM}{\mathcal{M}}
\newcommand{\cN}{\mathcal{N}}

\newcommand{\ot}{\hat \omega}
\newcommand{\Ac}{\check{\mathcal A}}
\newcommand{\Ah}{\hat{\mathcal A}}

\newcommand{\beq}{\begin{equation}}
\newcommand{\eeq}{\end{equation}}

\newcommand{\AdS}{{\text{AdS}}}
\newcommand{\GDiff}{{\text{GDiff}}}
\newcommand{\gdiff}{{\mathfrak{gdiff}}}

\newcommand{\aA}{A_i}

\newcommand{\MH}{\mathcal{A}_\text{H}}					

\title{Exactly marginal deformations from exceptional generalised geometry}

\author[a]{Anthony Ashmore,}
\author[b]{Maxime Gabella,}
\author[c]{Mariana Gra\~na,}
\author[d]{Michela Petrini}
\author[a]{and Daniel Waldram}

\affiliation[a]{Department of Physics, Imperial College London, London SW7 2AZ, UK}

\affiliation[b]{Institute for Advanced Study, Einstein Drive, Princeton, NJ 08540, USA}
\affiliation[c]{Institut de Physique Th\'eorique, CEA/Saclay, 91191 Gif-sur-Yvette, France}
\affiliation[d]{Sorbonne Universit\'e, UPMC Paris 05, UMR 7589, LPTHE, 75005 Paris, France}

\emailAdd{a.ashmore12@imperial.ac.uk}
\emailAdd{gabella@ias.edu}
\emailAdd{mariana.grana@cea.fr}
\emailAdd{petrini@lpthe.jussieu.fr}
\emailAdd{d.waldram@imperial.ac.uk}

\subheader{\textrm{Imperial/TP/16/DW/03 \\
		IPhT-t16/035}}
      
\abstract{We apply exceptional generalised geometry to the study of exactly marginal deformations of $\mathcal{N}=1$ SCFTs that are dual to generic AdS$_5$ flux backgrounds in type IIB or eleven-dimensional supergravity. In the gauge theory, marginal deformations are parametrised by the space of chiral primary operators of conformal dimension three, while exactly marginal deformations correspond to quotienting this space by the complexified global symmetry group. We show how the supergravity analysis gives a geometric interpretation of the gauge theory results. The marginal deformations arise from deformations of generalised structures that solve moment maps for the generalised diffeomorphism group and have the correct charge under the generalised Reeb vector, generating the R-symmetry. If this is the only symmetry of the background, all marginal deformations are exactly marginal. If the background possesses extra isometries, there are obstructions that come from fixed points of the moment maps. The exactly marginal deformations are then given by a further quotient by these extra isometries.
	
Our analysis holds for any $\mathcal{N}=2$ AdS$_5$ flux background. Focussing on the particular case of type IIB Sasaki--Einstein backgrounds we recover the result that marginal deformations correspond to perturbing the solution by three-form flux at first order. In various explicit examples, we show that our expression for the three-form flux matches those in the literature and the obstruction conditions match the one-loop beta functions of the dual SCFT.}

\begin{document} 

\maketitle

\section{Introduction}

The AdS/CFT correspondence allows the study of a wide class of superconformal field theories in four dimensions, many of which are realised as the world-volume theories of D3-branes at conical singularities of Calabi--Yau manifolds. Examples are ${\cal N}=4$ super Yang--Mills or the Klebanov--Witten model, which are obtained by putting D3-branes in flat space or at the tip of the cone over $\text{T}^{1,1}$ respectively.   

An interesting feature of $\cN=1$ SCFTs is that they can admit exactly marginal deformations, namely deformations that preserve supersymmetry and conformal invariance. A given $\cN=1$ SCFT can then be seen as a point on a ``conformal manifold'' in the space of operator couplings. The existence and dimension of the conformal manifold for a given theory can be determined using $\cN=1$ supersymmetry and renormalisation group arguments~\cite{LS95,Kol,GKSTW10,Kol2010}. For instance, ${\cal N}=4$ super Yang--Mills admits two exactly marginal deformations, the so-called $\beta$- and cubic deformations. What is more difficult to determine is the precise geometry of the conformal manifold.
 
Using AdS/CFT, the same questions can be asked by studying deformations of the supergravity background dual to the given SCFT. For ${\cal N}=4$ super Yang--Mills, the supergravity dual of the full set of marginal deformations is known only perturbatively. In~\cite{GPPZ98}, the first-order perturbation was identified with the three-form fluxes of type IIB, and the corresponding linearised solution was given in~\cite{GP01}. The second-order solution, including the back-reacted dilaton and metric, was constructed in~\cite{AKY02}, which also identified an obstruction to the third-order solution, corresponding to the vanishing of the gauge theory beta functions. This required considerable effort, and it seems far from promising to reconstruct the full solution from a perturbative analysis. On the other hand, using duality transformations, Lunin and Maldacena were able to build the full analytic supergravity dual of the $\beta$-deformation~\cite{LM05}. The same transformation applied to $\text{T}^{1,1}$ or $\text{Y}^{p,q}$ manifolds gives the gravity duals of the $\beta$-deformation of the Klebanov--Witten theory and more general $\mathcal{N}=1$ quiver gauge theories~\cite{LM05}. For the other marginal deformations of $\text{Y}^{p,q}$ models, the identification of the gravity modes dual to them can be found in~\cite{AP14} but no finite-deformation gravity solutions are known. 

The Lunin--Maldacena (LM) solution has a nice interpretation in generalised complex geometry~\cite{MPZ06, BFMMPZ08}, a formalism that allows one to geometrise the NS-NS sector of supergravity~\cite{Hitchin02,Gualtieri04}. One considers a generalisation of the tangent bundle of the internal manifold, given by the sum of the tangent and cotangent bundles. The structure group of this generalised tangent bundle is the continuous T-duality group $\Orth{d,d}$. The transformation that generates the LM solution is then identified as a bi-vector deformation inside $\Orth{d,d}$~\cite{MPZ06}. However, this is not the case for the other marginal deformation of $\cN=4$. In order to capture all exactly marginal deformations, one is tempted to look at the full U-duality group. This requires considering exceptional or $\Ex{d(d)}\times\mathbb{R}^{+}$ generalised geometry~\cite{Hull07,PW08}, where the U-duality groups appear as the structure groups of even larger extended tangent bundles.

The main motivation for this paper is to lay the foundations for applying exceptional generalised geometry to the study of exactly marginal deformations of a generic SCFT with a supergravity dual. As the first step of this programme we perform a linearised analysis of the exactly marginal deformations. To do this, we use the description of $\mathcal{N}=2$ AdS$_5$ backgrounds in terms of ``exceptional Sasaki--Einstein'' structures, given in~\cite{AW15b}. This is a generalisation of the conventional $G$-structure formalism where generalised structures are defined by generalised tensors that are invariant under some subgroup of $\Ex{d(d)}\times\mathbb{R}^{+}$. The relevant structures for AdS$_5$ compactifications are a hypermultiplet (or H) structure $J_\alpha$ and a vector-multiplet (or V) structure $K$. These structures are naturally associated with the hypermultiplet and vector-multiplet degrees of freedom of the five-dimensional gauged supergravity on AdS$_5$, hence their names~\cite{AW15}. Together they are invariant under a $\USp6$ subgroup of $\Ex{6(6)}\times\bbR^+$ and also admit a natural action of the $\USp2$ local symmetry of $\cN=2$ supergravity in five dimensions.\footnote{We use the standard nomenclature ${\mathcal N}=2$ to denote backgrounds with eight supercharges in five dimensions, even though this is the minimal amount of supersymmetry.} Although our specific examples will focus on type IIB geometries, the same analysis applies equally to generic $\cN=2$ AdS$_5$ solutions of type IIB or eleven-dimensional supergravity. 

This generalised geometric description of the internal geometry translates naturally to quantities in the dual field theory, which is particularly useful when analysing marginal deformations. Indeed, since hypermultiplets and vector multiplets of the gauged supergravity correspond to chiral and vector multiplets of the dual SCFT~\cite{Tachikawa06}, the deformations of the H and V structures map directly to superpotential and K\"ahler deformations of the dual SCFT. Using the properties of the $\cN=1$ superconformal algebra, Green et al.~\cite{GKSTW10} showed that marginal deformations can only be chiral operators of (superfield) dimension three and that the set of exactly marginal deformations is obtained by quotienting the space of marginal couplings by the complexified global symmetry group. The main result of this paper will be to reproduce these features from deformations of generic solutions on the supergravity side: the supersymmetric deformations must preserve the V structure but can deform the H structure. In addition, the exactly marginal deformations are a symplectic quotient of the marginal deformations by the isometry group of the internal manifold. This corresponds to the global symmetry group  of the dual field theory.

The paper is organized as follows: we begin in section 2 with a discussion of marginal deformations of $\mathcal{N}=1$ SCFTs focussing on a number of classic examples that are dual to AdS$_5 \times M$ type IIB backgrounds, where $M$ a Sasaki--Einstein manifold. In section 3, we review the reformulation of AdS$_5$ backgrounds in terms of exceptional generalised geometry \cite{AW15b,GN16}. We then describe how the moduli space of generalised structures appears and outline how this reproduces the findings of \cite{Kol,GKSTW10,Kol2010}. For concreteness, in section 4 we specialise to type IIB Sasaki--Einstein backgrounds. We find the explicit linearised supersymmetric deformations corresponding to the operators in the chiral ring, matching the Kaluza--Klein analysis of~\cite{EST13}, and recover the result that the supersymmetric deformations give rise to three-form flux perturbations~\cite{GP01}. In section 5, we give the explicit examples of $\text{S}^5$, $\text{T}^{1,1}$ and $\text{Y}^{p,q}$, and show that our expression for the three-form flux on $\text{S}^5$ matches the supergravity calculation of Aharony et al.~\cite{AKY02}, and reproduces the flux of the LM solution for generic Sasaki--Einstein manifolds. We conclude with some directions for future work in section~6.


\section{Marginal deformations of \texorpdfstring{$\cN=1$}{N=1} SCFTs}
\label{sec:scft}

Conformal field theories can be seen as fixed points of the renormalisation group flow where the beta functions for all couplings vanish. Generically, since there are as many beta functions as there are couplings, CFTs correspond to isolated points in the space of couplings. This is not the case for supersymmetric field theories, where non-renormalisation theorems force the beta functions for the gauge and superpotential couplings to be linear combinations of the anomalous dimensions of the fundamental fields~\cite{LS95}. If global symmetries are present before introducing the marginal deformations, the number of independent anomalous dimensions will be smaller than the number of couplings and not all beta functions will be independent. The theory then admits a manifold of conformal fixed points, $\mathcal{M}_\text{c}$. This is equivalent to saying that a given SCFT at a point $p\in \mathcal{M}_\text{c}$ admits exactly marginal deformations, namely deformations that preserve conformal invariance at the quantum level. The dimension of the conformal manifold is given by the difference between the number of classically marginal couplings and the number of independent beta functions. The two-point functions of the exactly marginal deformations at each point $p\in\mathcal{M}_\text{c}$ defines a natural metric on $\mathcal{M}_\text{c}$ called the Zamolodchikov metric. 

Recently, developing the argument in \cite{Kol}, the authors of \cite{GKSTW10} proposed an alternative method to analyse the $\mathcal{N}=1$ exactly marginal deformations of four-dimensional SCFTs, which does not use explicitly the beta functions for the superpotential couplings, but instead relies on the properties of the $\mathcal{N}=1$ algebra. Take a four-dimensional $\mathcal{N}=1$ SCFT at some point $p$ in the conformal manifold, and consider all possible marginal deformations. These are of two types: ``K\"ahler deformations'' which are perturbations of the form $\int \!\dd^4\theta\,V$ where $V$ is a real primary superfield of mass dimension $\Delta=2$, and ``superpotential'' deformations which have the form $\int \!\dd^2\theta\,\mathcal{O}$ where $\mathcal{O}$ is a chiral primary superfield with $\Delta=3$.\footnote{Here we give the mass dimension of the operator written as an $\mathcal{N}=1$ superfield. In component notation, in both cases the contribution to the Lagrangian has dimension $\Delta = 4$.} The results of~\cite{GKSTW10} are that:
\begin{itemize}
\item 
there are no marginal K\"ahler deformations since they
correspond to conserved currents;
\item 
there is generically a set of marginal superpotential deformations $\mathcal{O}_i$, with the generic deformation $W=h^i\mathcal{O}_i$ parametrised by a set of complex couplings $\{h^i\}$; 
\item 
if the undeformed theory has no global symmetries other than the $\Uni{1}_\text{R}$ R-symmetry, all marginal deformations are exactly marginal;
\item however if the original SCFT has a global symmetry $G$ that is broken by the generic deformation $W=h^i\mathcal{O}_i$, then the conformal manifold, near the original theory, is given by the quotient of the space of marginal couplings by the complexified broken global symmetry group
\beq \label{Mc}
{\cal M}_\text{c}= \frac{\{h^i \}}{G^{\mathbb C}} \, ,
\eeq
where ${\cal M}_\text{c}$ is K\"ahler manifold with the Zamolodchikov metric.
\end{itemize}
The reduction~\eqref{Mc} can be viewed as a symplectic quotient for the real group $G$, where setting the moment maps to zero corresponds to solving the beta function equations for the deformations. Note also that the vector space of couplings $h^i$ (modulo $G^{\mathbb C}$) parametrise the tangent space $T_p{\cal M}_\text{c}$ at the particular SCFT $p\in{\cal M}_\text{c}$, and so define local coordinates on the conformal manifold near $p$. Thus, as written~\eqref{Mc}, is only a local definition. 

More generally one can also consider operators $\mathcal{O}=A+ \theta\psi + \theta^2 F_A$ that are chiral primary superfields of any dimension, modulo the relations imposed by the F-terms of the SCFT. The lowest components $A$ form the chiral ring under multiplication $A''=AA'$ subject to the F-term relations, whereas the $\theta^2$-components satisfy $F_{A''}=AF_{A'}+A'F_A$, and hence transform as a derivation on the ring (specifically like a differential ``$\dd A$''). In what follows it will be useful to define the infinite-dimensional complex space of couplings $\{\gamma^i,\gamma'^i\}$ corresponding to deforming the Lagrangian by a term $\Delta=\gamma^i F_{A_i}+\gamma'^i A_i$ for generic chiral ring elements $A_i$ and $\theta^2$-components $F_{A_i}$. The $\gamma^i$ terms are supersymmetric, while the $\gamma'^i$ terms break supersymmetry, and generically neither are marginal. One of our results is that the supergravity analysis implies that there is a natural hyper-K\"ahler structure on this space, since the pair $(\gamma^i,\gamma'^i)$ arise from the scalar components of a hypermultiplet in the bulk AdS space. More precisely, if there is a global symmetry $G$, one naturally considers the space defined by the hyper-K\"ahler quotient\footnote{For more on this hyper-K\"ahler quotient see section~\ref{sec:exact}.}
\begin{equation} \label{Mtilde}
   \widetilde{\mathcal{M}} = \{ \gamma^i, \gamma'^i \} \qqquotient G \, . 
\end{equation}
The conformal manifold is then a finite-dimensional complex submanifold of $\widetilde{\mathcal{M}}$
\begin{equation}
   \mathcal{M}_\text{c} \subset \widetilde{\mathcal{M}} \, , 
\end{equation}
with the $\aA$ couplings $\gamma'^i$ set to zero and only the exactly marginal $\gamma^i$ coefficients (denoted $h_i$ above) non-zero.  

We now give three examples of SCFTs whose conformal manifolds have been analysed and whose gravity duals will be discussed in the rest of the paper.


\subsection{\texorpdfstring{$\cN=4$}{N=4} super Yang--Mills} 

The most studied example of a SCFT in four dimensions is $\mathcal{N}=4$ super Yang--Mills. The fields of the theory are -- besides gauge fields -- six scalars and four fermions, all in the adjoint representation of the gauge group $\SU{N}$ and transforming non-trivially under the $\SU{4}$ R-symmetry. In $\mathcal{N}=1$ notation, these fields arrange into a vector multiplet and three chiral superfields $\Phi^i$. The theory has a superpotential
\beq
\label{N4superpot}
W_{\mathcal{N}=4} = \tfrac{1}{6} h \epsilon_{ijk}\tr(\Phi^{i}\Phi^{j}\Phi^{k}) \,,
\eeq
which is antisymmetric in the fields, and the coupling is fixed by $\mathcal{N}=4$ supersymmetry to be equal to the gauge coupling, $h=\tau$. In this notation, only the $\SU{3} \times \Uni{1}$ subgroup of the R-symmetry is manifest.

The marginal deformations compatible with $\mathcal{N}=1$ supersymmetry are given by the chiral operators
\begin{equation}
\label{WN4def}
W=\tfrac{1}{6} h \epsilon_{ijk}\tr(\Phi^{i}\Phi^{j}\Phi^{k})+\tfrac{1}{6}f_{ijk}\tr(\Phi^{i}\Phi^{j}\Phi^{k})\,,
\end{equation}
where $f_{ijk}$ is a complex symmetric tensor of $\SU{3}$ and  $h$ is a priori different from the gauge coupling $\tau$. In all there are eleven complex marginal deformations. The superpotential \eqref{WN4def} breaks the global $\SU{3}$ symmetry, leaving the $\Uni{1}_\text{R}$ symmetry of $\mathcal{N}=1$ theories. Therefore, the conformal manifold is
\begin{equation}
\label{N4quotient}
\mathcal{M}_{\text{c}}= \frac{\{h, f_{ijk} \}}{\SU{3}^{\bbC}}  \,,
\end{equation}
with complex dimension $\dim(\mathcal{M}_{\text{c}})=11-8= 1 + 2$. 
The first deformation is an $\SU{4}$ singlet corresponding to changing both $\tau$ and $h$, the other two are true superpotential deformations.

The same conclusions can be reached by studying the beta functions of the deformed theory \cite{LS95, AKY02}. One can show that the beta function equations for the gauge coupling and the superpotential deformations are proportional to the matrix of anomalous dimensions. At one loop, this (or more precisely its traceless part) is
\begin{equation}\label{8conds}
\gamma_{i}{}^{j}=\frac{N^{2}-4}{64N\pi^{2}}(f_{ikl}\bar{f}^{jkl}-\tfrac{1}{3}\delta_{i}{}^{j}f_{klm}\bar{f}^{klm})=0 \, ,
\end{equation}
corresponding to the $\SU{3}$ moment maps, when we view~\eqref{N4quotient} as a symplectic quotient. This equation imposes eight real conditions on $f_{ijk}$. One can remove another eight real degrees of freedom using an $\SU{3}$ rotation of the fields $\Phi^i$. Together, these reduce the superpotential deformation to~\cite{LS95}
\beq
\label{WN4defbis}
W=\tfrac{1}{6} h \epsilon_{ijk}\tr(\Phi^{i}\Phi^{j}\Phi^{k})+ f_\beta \tr (\Phi^{1}\Phi^{2}\Phi^{3}+\Phi^{3}\Phi^{2}\Phi^{1})+ f_\lambda \tr\bigl((\Phi^1)^3+ (\Phi^2)^3 + (\Phi^3)^3 \bigr)  \,.
\eeq
The coupling $f_\beta$ is the so-called $\beta$-deformation,\footnote{This term can also be written as $ \tr (e^{ \ii \pi \beta} \Phi^{1}\Phi^{2}\Phi^{3} - e^{- \ii \pi \beta}  \Phi^{3}\Phi^{2}\Phi^{1})$ where $\beta$ is complex~\cite{LM05}.} and $f_\lambda$ is often called the cubic deformation. As mentioned above, the first term in this expression is to be interpreted 
as changing $h$ and $\tau$ together. 

One can go beyond the one-loop analysis. The deformed theory has a discrete $\mathbb{Z}_3 \times \mathbb{Z}_3$ symmetry, which forces the matrix of anomalous dimensions of the $\Phi^i$ to be proportional to the identity. One can then show that the beta function condition (at all loops) reduces to just one equation, thus again giving a three-dimensional manifold of exactly marginal deformations. Since this will be relevant for the gravity dual, we stress that the only obstruction to having exactly marginal deformations is the one-loop constraint \eqref{8conds}.


\subsection{Klebanov--Witten theory}

The Klebanov--Witten theory is the four-dimensional SCFT that corresponds to the world-volume theory of $N$ D3-branes at the conifold singularity~\cite{KW98}. This is an $\cN=1$ $\SU{N} \times \SU{N}$ gauge theory with two sets of bi-fundamental chiral fields $A_i$ and $B_i$ ($i=1,2$) transforming in the $(\rep{N}, \rep{\overline{N}})$ and $(\rep{\overline{N}}, \rep{N})$ respectively. The superpotential is
\beq
\label{WKW} 
W  = h \epsilon^{\alpha \beta}  \epsilon^{\dot{\alpha} \dot{\beta}}  \tr (A_\alpha B_{\dot{\alpha}} A_\beta B_{\dot{\beta}})  \, ,
\eeq
and preserves an $\SU{2} \times \SU{2} \times \Uni{1}_\text{R}$ global symmetry, under which the chiral fields transform as $(\rep{2},\rep{1},\rep{1/2})$ and $(\rep{1},\rep{2},\rep{1/2})$ respectively. The R-charges of the fields $A_i$ and $B_i$ are such that the superpotential has the standard charge $+2$. The superpotential is not renormalisable, suggesting that the theory corresponds to an IR fixed point of an RG flow. Indeed, one can show that this theory appears as the IR fixed point of the RG flow generated by giving mass to the adjoint chiral multiplet in the $\mathbb{Z}_2$ orbifold of $\cN =4$ super Yang--Mills~\cite{KW98}.

Classically, the marginal deformations of the KW theory are given by the following chiral operators
\begin{equation} \label{WKWdef} 
\begin{split}
W &=  h \epsilon^{\alpha \beta}  \epsilon^{\dot{\alpha} \dot{\beta}}  \tr (A_\alpha B_{\dot{\alpha}} A_\beta B_{\dot{\beta}}) +
f^{\alpha \beta , \dot{\alpha} \dot{\beta}} \tr (A_\alpha B_{\dot{\alpha}} A_\beta B_{\dot{\beta}}) \\
&\eqspace + \tau \bigl[   \tr (W_1 W_1) -  \tr (W_2 W_2) \bigr]  \, ,
\end{split}
\end{equation}
where the tensor $f^{\alpha \beta , \dot{\alpha} \dot{\beta}} $ is symmetric in the indices $\alpha\beta$ and $\dot{\alpha}\dot{\beta}$, and therefore transforms in the $(\rep{3},\rep{3})$ of the $ \SU{2} \times \SU{2} $ global symmetry group. The deformation $\tau$  does not break the global symmetry of the theory and corresponds to a shift in the difference of the gauge couplings $(1/g^2_{1} -1/g^2_{2})$.

The exactly marginal deformations of the KW theory were found in~\cite{BH05}. Only three components of the $f^{\alpha \beta , \dot{\alpha} \dot{\beta}}$ term are exactly marginal, so we have five exactly marginal deformations in total. This is in agreement with the dimension of the conformal manifold, given by
\beq
\mathcal{M}_{\text{c}}=\frac{ \{h,  f^{\alpha \beta ,  \dot{\alpha} \dot{\beta}} , \tau \} }{( \SU 2  \times  \SU 2)^{\bbC}}  \, .
\eeq

One reaches the same conclusions by studying the beta functions of the deformed theory \cite{KW98}. These are equivalent to the $ \SU{2} \times \SU{2} $ moment maps, which take the form
\begin{equation} \label{betaKW}
\begin{split}
\gamma^{\alpha}{}_{\beta}=f^{\alpha\gamma\dot{\alpha}\dot{\beta}} \bar{f}_{\beta\gamma\dot{\alpha}\dot{\beta}} - \tfrac{1}{2}\delta_{\phantom{\alpha}\beta}^{\alpha}f^{\tau\gamma\dot{\alpha}\dot{\beta}}\bar{f}_{\tau\gamma\dot{\alpha}\dot{\beta}} & =0 \,, \\
\gamma^{\dot{\alpha}}{}_{\dot{\beta}}=f^{\alpha\beta\dot{\alpha}\dot{\gamma}} \bar{f}_{\alpha\beta\dot{\beta}\dot{\gamma}}-\tfrac{1}{2}\delta_{\phantom{\alpha}\dot{\beta}}^{\dot{\alpha}}f^{\alpha\beta\dot{\tau}\dot{\gamma}}\bar{f}_{\alpha\beta\dot{\tau}\dot{\gamma}} & =0 \,.
\end{split}
\end{equation}
These remove six real degrees of freedom. We can also redefine the couplings using the $\SU{2}\times\SU{2}$ symmetry to remove another six real degrees of freedom, leaving three complex parameters. The exactly marginal deformations are then given by
\beq
\begin{split} \label{WemKW}
W &=  h \epsilon^{\alpha \beta}  \epsilon^{\dot{\alpha} \dot{\beta}}  \tr (A_\alpha B_{\dot{\alpha}} A_\beta B_{\dot{\beta}})  + \tau \bigl[   \tr (W_1 W_1) -  \tr (W_2 W_2) \bigr] \\
&\eqspace   + f_\beta ( A_1 B_{\dot{1}} A_2 B_{\dot{2}} + A_1 B_{\dot{2}}  A_2 B_{\dot{1}}  ) + f_{2} ( A_1 B_{\dot{1}}  A_1 B_{\dot{1}}  + A_2 B_{\dot{2}}  A_2 B_{\dot{2}}  ) \\
&\eqspace   + f_{3} ( A_1 B_{\dot{2}}  A_1 B_{\dot{2}}  + A_2 B_{\dot{1}}  A_2 B_{\dot{1}}  ) \, .
\end{split}
\end{equation}
The deformation parametrised by $f_\beta$ is the $\beta$-deformation for the KW theory, since it is the deformation that preserves the Cartan subgroup of the global symmetry group ($\Uni{1}\times \Uni{1}$ in this case). 

\subsection{\texorpdfstring{Y$^{p,q}$}{Y(p,q)} gauge theories} \label{sec:GTYpq}

The KW theory is the simplest example of an $\cN =1$ quiver gauge theory in four dimensions. A particularly interesting class of these theories  arise as world-volume theories of D3-branes probing a Calabi--Yau three-fold with a toric singularity, where the singular Calabi--Yau spaces are cones over the infinite family of Sasaki--Einstein $\text{Y}^{p,q}$ manifolds~\cite{GMSW04,GMSW04b}.\footnote{The integer numbers $p$ and $q$ satisfy $0\le q \le p$. Note that $\text{Y}^{1,0}=\text{T}^{1,1}$, the five-dimensional manifold in the KW theory.} These theories have rather unusual properties, such as the possibility of irrational R-charges. The field theories dual to the infinite family of geometries were constructed in \cite{BFHMS05}, which we review quickly.

The properties of the dual field theories can be read off from the associated quiver. The fields theories have $2p$ gauge groups with $4p+2q$ bi-fundamental fields. Besides the $\Uni{1}_\text{R}$, they have an $\SU{2}\times\Uni{1}_\text{F}$ global symmetry.  The $4p+2q$ fields split into doublets and singlets under $\SU{2}$: $p$ doublets labelled $U$, $q$ doublets labelled $V$, $p-q$ singlets labelled $Z$ and $p+q$  singlets labelled $Y$. The general superpotential is
\begin{equation}
W = h \epsilon_{\alpha\beta}\Biggl( \sum_{k=1}^q (U_{k}^{\alpha} V_{k}^{\beta} Y_{2k-1} + V_{k}^{\alpha} U_{k+1}^{\beta} Y_{2k})+\sum_{j=q+1}^p Z_{j} U_{j+1}^{\alpha} Y_{2j-1} U_{j}^{\beta} \Biggr) \, ,
\end{equation}
where the $\alpha$ and $\beta$ indices label the global $\SU{2}$. 
 The R-charges of the fields are
\begin{equation}
\begin{split}
r_U &= \tfrac{2}{3}pq^{-2}\bigl(2p-(4p^2-3q^2)^{1/2}\bigr) \, , \\
r_V &= \tfrac{1}{3}q^{-1}\bigl(3q-2p+(4p^2-3q^2)^{1/2}\bigr) \, , \\
r_Y &= \tfrac{1}{3}q^{-2}\bigl(-4p^2+3q^2+2pq+(2p-q)(4p^2-3q^2)^{1/2}\bigr) \, , \\
r_Z &= \tfrac{1}{3}q^{-2}\bigl(-4p^2+3q^2-2pq+(2p+q)(4p^2-3q^2)^{1/2}\bigr) \, ,
\end{split}
\end{equation}
while their charges under the additional $\Uni{1}_\text{F}$ symmetry are respectively $0$, $1$, $-1$ and $1$.

The marginal deformations of these theories are given by~\cite{BH05}
\begin{equation} \label{WYpq}
W= (h \epsilon_{\alpha\beta} + f_{\alpha\beta}) \sum \Bigl(U_{k}^{\alpha}V_{k}^{\beta}Y_{2k-1}+V_{k}^{\alpha} U_{k+1}^{\beta} Y_{2k} + Z_j U^{\alpha}_{j+1} Y_{2j-1} \, , U_j^{\beta} \Bigr)+  \tau \, {\cal O}_{\rm{gauge}}\, ,
\end{equation}
where $f_{\alpha \beta}$ is symmetric and ${\cal O}_{\text{gauge}}$ is an operator involving differences of gauge couplings. Note that $W$ preserves $\Uni{1}_\text{F}$, but the $f_{\alpha\beta}$ terms break the $\SU{2}$ to $\Uni{1}$. The $\SU{2}$ moment maps giving the beta functions are
\begin{equation} \label{betaYpq}
 \epsilon_{abc} f^b \bar f^c  =0 \, ,
\end{equation}
where $f_{\alpha\beta}=f^a (\sigma_a)_{\alpha \beta}$, which has the solution $f^a=r^a \ee^{\ii \phi}$. Modding out by the SU(2) action leaves a single  deformation that is exactly marginal, namely the analogue of the $\beta$-deformation for the $\text{Y}^{p,q}$ theories. As mentioned previously, the $\beta$-deformation breaks the global symmetry to its Cartan generators. Thus one can take $f^3$ non-zero, or equivalently
\beq
f_{11}=-f_{22}\equiv f_{\beta} \ .
\eeq
Note that the counting is in agreement with the dimension of the conformal manifold, given by
\begin{equation}
\mathcal{M}_{\text{c}} = \frac{\{h,f_{{\alpha}\beta}, \tau\}}{\SU{2}^\mathbb{C}} =\{h, f_{\beta},\tau \} \, .
\end{equation}
Naively the quotient gives the wrong counting. However $f_{\alpha\beta}$ does not completely break $\SU{2}$ but instead preserves a $\Uni{1}$, meaning that the quotient removes only two complex degrees of freedom.


\section{Deformations from exceptional generalised geometry}
\label{sec:sugraduals}
 
According to AdS/CFT, the supergravity dual of a given conformal field theory in four dimensions is a geometry of the form $\text{AdS}_5 \times M$, where the $\text{AdS}_5$ factor reflects the conformal invariance of the theory. The duals of exactly marginal deformations that preserve $\cN=1$ supersymmetry are expected to be of the same form, but with a different geometry on the internal manifold. Generically, the solution will also have non-trivial fluxes and dilaton, if present. These solutions should be parametrically connected to the undeformed solution, so that the moduli space of exactly marginal deformations of the gauge theory is mapped to the moduli space of $\text{AdS}_5$ vacua. 

Finding the full supergravity duals of exactly marginal deformations is not an easy task; few exact solutions are known, and those that are were found using solution-generating techniques based on dualities~\cite{LM05}. The idea of this paper is to exploit as much of the symmetry structure of the supergravity as possible to look for the generic exactly marginal deformations. This is most naturally done in the context of exceptional generalised geometry, where by considering an extended tangent bundle that includes vectors, one-forms and higher-rank forms, one finds an enhanced $\Ex{d(d)}\times\bbR^+$ structure group and the bosonic fields are unified into a generalised metric. 

In this section, we outline the general results applicable to arbitrary $\text{AdS}_5$ supergravity backgrounds, whether constructed from type II or eleven-dimensional supergravity. In particular, we find the supergravity dual of the field theory results of~\cite{GKSTW10}. In the following section, we discuss the specific case of type IIB compactifications on Sasaki--Einstein manifolds giving considerably more detail. 

\subsection{Generalised structures and deformations}
\label{K_left_inv}

Consider a generic supersymmetric solution of the form $\AdS_5\times M$, where $M$ can be either five- or six-dimensional depending on whether we are compactifying type II or eleven-dimensional supergravity. We allow all fluxes that preserve the symmetry of $\AdS_5$.

We are looking for the duals of $\cN=1$ SCFTs in four dimensions and so the dual supergravity backgrounds preserve eight supercharges, that is $\cN=2$ in five dimensions. A background preserving eight supercharges is completely determined by specifying a pair of generalised structures~\cite{AW15}: a ``vector-multiplet structure'' $K$ and a ``hypermultiplet structure'' $J_\alpha$, a triplet of objects labelled by $\alpha=1,2,3$. Each structure is constructed as a combination of tensors on $M$ built from bilinears of Killing spinors~\cite{GN16}, but for the moment the details are irrelevant. One should think of them as defining a generalisation of the Sasaki--Einstein structure in type IIB to a generic $\AdS_5$ flux background in type II or M-theory. 

Supersymmetry implies that the structures $K$ and $J_\alpha$ satisfy three differential conditions~\cite{AW15b,GN16}. The two of particular relevance to us take the form
\begin{align}
  \label{mmconditions} 
   \mu_\alpha(V) &= \lambda_\alpha \int c(K,K,V)   
      \qquad \forall\, V \,, \\
  \label{genLieJa}
  \Dorf_{K}  J_\alpha &= \epsilon_{\alpha\beta\gamma} \lambda_\beta J_\gamma \,, 
\end{align}
where the triplet of functions $\mu_\alpha(V)$ are defined to be 
\begin{equation}
   \mu_\alpha(V) \coloneqq - \tfrac{1}{2}\epsilon_{\alpha\beta\gamma}\int 
      \tr (J_\beta\,\Dorf_V J_\gamma) .
\end{equation}
The third condition is 
\beq \label{LKK}
\Dorf_K K=0 \ .
\eeq 
The constants $\lambda_\alpha$ are related to the $\text{AdS}_5$ cosmological constant and can always be fixed to 
\beq \label{lambdaa}
\lambda_1 = \lambda_2 = 0 \,, \qqq \lambda_3 = 3 \,.
\eeq
Again the details are not important here, but for completeness note that $c(K,K,V)$ is the $\Ex{6(6)}$ cubic invariant (see~\eqref{eq:IIB_cubic}) while the symbol $\Dorf$ denotes the Dorfman or generalised Lie derivative (see~\eqref{IIB_twisted_Dorf}), which generates the group of generalised diffeomorphisms $\GDiff$, namely the combination of diffeomorphisms and gauge transformations of all the flux potentials. In particular one can show that $K$ is a ``generalised Killing vector'', that is $\Dorf_K$ generates a generalised diffeomorphism that leaves the solution invariant, and this symmetry corresponds to the R-symmetry of the SCFT. In analogy to the Sasaki--Einstein case, we sometimes refer to $K$ as the ``generalised Reeb vector''. In addition, the functions $\mu_\alpha$ have an interpretation as a triplet of moment maps for the group of generalised diffeomorphisms acting on the space of $J_\alpha$ structures. As such we will often refer to~\eqref{mmconditions} as the moment map conditions. 

To find the marginal deformations of the $\mathcal{N}=1$ SCFT we need to consider perturbations of the structures $K$ and $J_\alpha$ that satisfy the supersymmetry conditions, expanded to first order in the perturbation. These are of two types,\footnote{There is actually a third type where both $\delta J_\alpha\neq 0$ and $\delta K\neq0$, but in this case none of the supergravity fields are perturbed: instead it corresponds to a deformation of the Killing spinors, implying the background admits more than eight supersymmetries. For this reason it will not interest us here.} which correspond to the two types of deformation in the SCFT
\begin{equation*}
\begin{aligned}
   \delta K &\neq 0 \, , &  \delta J_\alpha &= 0 
      & :& & &\text{K\"ahler deformations,} \\ 
   \delta K &= 0 \, , & \delta J_\alpha &\neq 0 
      & :& & &\text{superpotential deformations.} 
\end{aligned}
\end{equation*}
The easiest way to justify this identification is to note that, from the point of view of five-dimensional supergravity, fluctuations of $K$ live in vector multiplets and those of $J_\alpha$ live in hypermultiplets. According to the AdS/CFT dictionary, vector multiplets and hypermultiplets correspond to real primary superfields and chiral primary superfields in the SCFT~\cite{Tachikawa06}. 

Let us first consider the K\"ahler deformations, where we hold $J_\alpha$ fixed and deform $K$. Looking at the moment maps~\eqref{mmconditions}, we see the left-hand side depends only on $J_\alpha$ and so does not change, but the right-hand side can vary, thus we have 
\begin{equation}
   \int c(K,\delta K,V) = 0  \qquad \forall V \,. 
\end{equation}
The $K$ tensor is invariant under a $\Fx{4(4)}\subset\Ex{6(6)}$ subgroup. A simple argument using the decomposition into $\Fx{4(4)}$ representations, then implies that $\delta K=0$. This matches the field theory analysis that there are no deformations of K\"ahler type. 

For the superpotential deformations we can solve~\eqref{mmconditions} and~\eqref{genLieJa} to first order in $\delta J_\alpha$. We can do this in two steps. First we solve the linearised moment map conditions~\eqref{mmconditions}. This gives an infinite number of solutions which correspond to $\theta^2$-components and fields in the chiral ring of the dual gauge theory; generically these are not marginal. Imposing the first-order generalised Lie derivative condition~\eqref{genLieJa} will select a finite number of these modes that are massless in $\text{AdS}_5$ and correspond to the actual marginal deformations. 

\subsection{Exactly marginal deformations and fixed points}
\label{sec:exact}

We now turn to how the supergravity structure encodes the SCFT result that all marginal deformations are exactly marginal unless there is an additional global symmetry group $G$. The key point, as we will see, is that the differential conditions~\eqref{mmconditions} appear as moment maps for the generalised diffeomorphisms.

A priori, to see if the marginal deformations are exactly marginal one needs to satisfy the equations~\eqref{mmconditions} and~\eqref{genLieJa} not just to first order, but to all orders in the deformation. In general this is a complicated problem: typically there can be obstructions at higher order that mean not all marginal deformations are actually exactly marginal. For example, a detailed discussion of deformations of $\mathcal{N}=4$ up to third order is given in~\cite{AKY02}.

However, viewing the conditions~\eqref{mmconditions} as a triplet of moment maps provides an elegant supergravity dual of the field theory result that does not require detailed case-by-case calculations. The generic situation is discussed in some detail in~\cite{AW15b}, which we now review. Moment maps arise when there is a group action preserving a symplectic or hyper-K\"ahler structure. Here the $\mu_\alpha$ correspond to the action of generalised diffeomorphisms -- that is conventional diffeomorphisms and/or a form-field gauge transformations -- acting on the structure $J_\alpha$. Thus to get physically distinct solutions we need to satisfy the moment map conditions~\eqref{mmconditions} and then identify solutions that are related by a generalised diffeomorphisms. Formally this defines a subspace of hypermultiplet structures 
\begin{equation} \label{tildeM}
   \widetilde{\mathcal{M}} 
      = \frac{\{ J_\alpha : \mu_\alpha=\lambda_\alpha\gamma \}}{\GDiff_K} \, ,
\end{equation}
where $\gamma$ is the function 
\begin{equation}\label{alphadef}
   \gamma(V) = \int c(K,K,V) \, ,
\end{equation}
and $\GDiff_K$ is the subgroup of generalised diffeomorphisms that leave $K$ invariant. (We are considering the moduli space of solutions for $J_\alpha$ for fixed $K$.) By construction~\eqref{tildeM} defines a hyper-K\"ahler quotient and hence $\widetilde{\mathcal{M}}$ is hyper-K\"ahler. The condition~\eqref{genLieJa} then defines a K\"ahler subspace $\mathcal{M}_\text{c}\subset\widetilde{\mathcal{M}}$ (see~\cite{AW15b}). We can also consider first imposing~\eqref{genLieJa} and then the moment maps~\eqref{mmconditions}. Let~$\MH^K$ be the space of H-structures $J_\alpha(x)$ for fixed $K$. Imposing~\eqref{genLieJa} defines a K\"ahler subspace $\mathcal{N}_{\text{H}}\subset\MH^K$. The moment map conditions then take a symplectic quotient of  $\mathcal{N}_{\text{H}}$ rather than a hyper-K\"ahler quotient. We then have the following picture~\cite{AW15b}
\begin{equation}
\label{eq:quotients}
\begin{tikzcd}[row sep=large, column sep=huge]
\MH^K \arrow{r}{\eqref{genLieJa}}
\arrow{d}[swap]{\text{HK quotient~\eqref{mmconditions}}}
& \mathcal{N}_{\text{H}} \arrow{d}{\text{sympl.~quotient~\eqref{mmconditions}}} \\
\widetilde{\mathcal{M}} \arrow{r}{\eqref{genLieJa}} & \mathcal{M}_\text{c} 
\end{tikzcd}
\end{equation}

A nice property of moment map constructions is that generically there are no obstructions to the linearised problem: every first-order deformation around a given point $p\in\widetilde{\mathcal{M}}$ in the hyper-K\"ahler quotient (or alternatively $p\in\mathcal{M}_\text{c}$ for the symplectic quotient) can be extended to an all-order solution. The only way this fails is if the symmetry group at $p$ defining the moment map has fixed points. In our context this means there are generalised diffeomorphisms that leave the particular $J_\alpha$ and $K$ structures invariant, that is one can find $V$ such that the generalised Lie derivatives vanish
\begin{equation}
   \Dorf_V J_\alpha = \Dorf_V K = 0 \,. 
\end{equation}
These $V$ generate isometries of the background (beyond the $\Uni{1}_\text{R}$ R-symmetry), corresponding to the global symmetry group $G$ of the dual field theory. In other words, the vector component of $V$ is a Killing vector.\footnote{For example, for $M=\text{S}^5$ the isometry group is $\SO6 \simeq \SU4 \supset \Uni{1}_\text{R} \times \SU3$, so $V$ would give the Killing vectors that generate $\SU3$.} Thus we directly derive the result that every marginal deformation is exactly marginal in the absence of global symmetries. 

Suppose now that the global symmetry group $G$ is non-trivial. By construction, those $V$ that generate $G$ fall out of the linearised moment map conditions. Thus to solve the full non-linear problem, one must somehow impose these additional conditions. It is a standard result in symplectic (or hyper-K\"ahler) quotients that the missing equations correspond to a quotient by the global group $G$ on the space of linearised solutions. Suppose $\{\gamma^i,\gamma'^i\}$ are coordinates on the space of linearised deformations, corresponding to couplings of operators $F_{A_i}$ and $A_i$. Imposing~\eqref{genLieJa} then restricts to the marginal operators $\{h_i\}\subset \{\gamma^i,\gamma'^i\}$. By construction, there is a flat hyper-K\"ahler metric on $\{\gamma^i,\gamma'^i\}$ and a flat K\"ahler metric on $\{h_i\}$. In addition there is a linear action of $G$ on each space that preserves these structures. The origin is a fixed point of $G$ corresponding to the fact that we are expanding about a solution with a global symmetry. The moduli space of finite deformations then corresponds to a quotient of each space by $G$ (at least in the neighbourhood of the original solution). Thus we have 
\begin{equation}
\label{eq:final-quotients}
   \begin{tikzcd}[row sep=large, column sep=huge]
      \{\gamma^i,\gamma'^i\} \arrow{r}{\eqref{genLieJa}}
            \arrow{d}[swap]{\text{HK quotient by $G$}} 
         &  \{h_i\} 
            \arrow{d}{\text{sympl.~quotient by $G$}} \\
      \widetilde{\mathcal{M}} \arrow{r}{\eqref{genLieJa}} & \mathcal{M}_\text{c} 
   \end{tikzcd}  
\end{equation}
This structure is discussed in little more detail in section~\ref{sec:obstructions}. We see that we directly recover the field theory result~\eqref{Mc} that the conformal manifold is given by $\mathcal{M}_{\text{c}}=\{h_i\}\qquotient G=\{h_i\}\quotient G^\bbC$.

Note that interpreting the supersymmetry conditions in terms of moments maps
nicely mirrors the field theory analysis of the moduli space of marginal deformations.  Indeed imposing \eqref{genLieJa} and solving the linearised moment maps~\eqref{mmconditions} is  equivalent to restricting to chiral operators of dimension three that satisfy the F-term conditions. The further symplectic quotient by the isometry group $G$ then corresponds to imposing the D-term constraints and modding out by gauge transformations.


\section{The case of D3-branes at conical singularities}
\label{sec:SE}

The results summarized in the previous section are completely general and apply to any $\AdS_5$ flux background. To make the discussion more concrete we will focus on deformations of $\mathcal{N}=1$ SCFTs that are realised on the world-volume of D3-branes at the tip of a Calabi--Yau cone over a Sasaki--Einstein (SE) manifold $M$.  
 
Before turning to the generalized geometric description of the supergravity duals, we present their description in terms of ``conventional" geometry.

\subsection{The undeformed Sasaki--Einstein solution}

In the ten-dimensional type IIB solution dual to the undeformed SCFT, the metric takes the form\footnote{In these conventions the radius of $\text{AdS}_{5}$ is $R=1$, so the cosmological constant is $\Lambda = -6$.}
\begin{equation}
\begin{split}
\dd s_{10}^{2} & =e^{2A}\dd s^{2}(\mathbb{R}^{1,3})+e^{-2A}\dd s^{2}(\text{CY})\label{10dmet} \\
 & =r^{2}\eta_{\mu\nu}\dd x^{\mu}\dd x^{\nu}+\frac{1}{r^{2}}\bigl(\dd r^{2}+r^{2}\dd s^{2}(\text{SE})\bigr) \\
 & =\dd s^{2}(\text{AdS}_5)+\dd s^{2}(\text{SE}) \,,
\end{split}
\end{equation}
where the radial direction of the cone together with the four-dimensional warped space form  $\text{AdS}_5$. In the second and third line we have used the explicit form of the warp factor for $\text{AdS}_{5}$, $e^{A}=r$. The solution has constant dilaton, $e^{\phi}=1$, and five-form flux given by
\beq \label{F5}
F_5 = 4( \vol_\text{AdS}  + \vol_5) \, ,
\eeq
where $\vol_{5}$ is the volume form on $M$.

The metric on the SE manifold locally takes the form 
\begin{equation} 
\label{SEKE}
\dd s^2_{\rm{SE}} = \sigma^2 + \dd s^2_{\rm{KE}} \, ,
\end{equation}
where $\sigma$ is called the contact form and the four-dimensional metric is K\"ahler--Einstein (KE), with symplectic two-form given by   
\beq
\label{eq:dsigma}
\omega = \tfrac{1}{2} \dd \sigma \, .
\eeq
There is also a holomorphic (2,0)-form $\Omega$, compatible with $\omega$
\beq
\label{eq:vol4}
\omega \wedge \Omega =0 \,, \qqq \omega \wedge \omega = \tfrac{1}{2} \Omega \wedge \bar \Omega \,,
\eeq
satisfying 
\beq 
\label{eq:dOmega}
 \dd\Omega = 3\ii \sigma \wedge \Omega \,.
\eeq
The five-dimensional volume form is then
\beq\label{eq:vol5}
 \vol_5 = -\tfrac{1}{2} \sigma \wedge \omega \wedge \omega \,.
\eeq
The forms $\sigma$, $\Omega$ and $\omega$ define an $\SU2$ structure on the Sasaki--Einstein manifold. 

It will be useful in what follows to introduce a $(2,0)$ bi-vector $\alpha$ that is obtained from $\bar{\Omega}$ by raising its indices with the metric. Then the complex structure $I$ for the K\"ahler--Einstein metric can be written as 
\begin{equation}
I^{m}_{\phantom{m}n}=-\omega^{m}_{\phantom{m}n}=\tfrac{1}{4}\ii({\alpha}^{mp}\Omega_{np}-\bar{\alpha}^{mp}\bar{\Omega}_{np}) \,.
\end{equation}

The R-symmetry of the field theory is realised in the dual geometry by the Reeb vector field $\xi$ satisfying 
\beq\label{eq:asigma}
\imath_{\xi} \sigma = 1\, , \qqq \imath_{\xi} \dd\sigma =0 \,.
\eeq
Locally we can introduce a coordinate $\psi$ such that
\beq \label{sigma}
\sigma=\tfrac{1}{3}(\dd \psi + \eta) \ , \qqq \xi=3\partial_{\psi} \,.
\eeq
If a tensor  $X$ satisfies $\mathcal{L}_{\xi}X=\ii q X$, we say it has charge $q$ under the action of the Reeb vector. The objects defining the $\SU{2}$ structure on $M$ have definite charge
\beq \label{scalingomega}
\mathcal{L}_{\xi}\sigma=\mathcal{L}_{\xi}\omega=\mathcal{L}_{\xi}I=0 \,,\qqq\mathcal{L}_{\xi}\Omega=3i\Omega \,.
\eeq
The R-charge $r$ is related to $q$ by $q=3r/2$. For example, $\Omega$ is charge $+3$ under the Reeb vector and has R-charge $+2$.

The contact and K\"ahler structures allow a decomposition of the exterior derivative as
\beq 
\label{decompositiond}
\dd = \del + \bar \del + \sigma \wedge \cL_\xi \,,
\eeq
where $\bar{\partial}$ is the tangential Cauchy-Riemann operator, which satisfies~\cite{KR65}
\begin{equation}
\bar{\partial}^2=\partial^2=0 \,,\qqq \partial\bar{\partial}+\bar{\partial}\partial=-2\omega\wedge\mathcal{L}_{\xi} \,.
\end{equation}
For calculations, it is useful to introduce a frame such that the complex, symplectic and contact structure have the following form
\begin{equation} \label{frame}
\begin{split}\Omega & =(e^{2}+\ii e^{5})\wedge(e^{4}+\ii e^{3}) \,,\\
\omega & =e^{2}\wedge e^{5}+e^{4}\wedge e^{3} \,,\\
\sigma & =e^{1} \,.
\end{split}
\end{equation}
In terms of the dual frame the bi-vector $\alpha$  is
\beq
\alpha = (\hat{e}_{2}-\ii \hat{e}_{5})\wedge(\hat{e}_{4}-\ii \hat{e}_{3}) \,.\\
\eeq

If the SE manifold is ``regular" the Reeb vector defines a $\Uni{1}$ fibration over a K\"ahler--Einstein base. This is the case for $\text{S}^5$ and $\text{T}^{1,1}$, dual to $\cN=4$ SYM and the $\cN =1$ KW theory, where the base manifolds are respectively  $\mathbb{CP}^2$ and $\mathbb{CP}^1 \times \mathbb{CP}^1$. The $\text{Y}^{p,q}$ spaces are generically not fibrations. 


\subsection{Embedding in exceptional generalised geometry}
\label{sec:JK}

In this section we review the description of supersymmetric $\AdS_5 \times M$ solutions in $\Ex{6(6)}\times\mathbb{R}^{+}$ generalised geometry following~\cite{AW15,AW15b}. Although we will focus on type IIB for definiteness, we stress that the construction is equally applicable to solutions of eleven-dimensional supergravity. In particular, details of the embedding of the generic M-theory $\AdS_5$ solution of~\cite{GMSW04c} into $\Ex{6(6)}\times\mathbb{R}^{+}$ generalised geometry are given in~\cite{AW15b}. 

The idea of exceptional generalised geometry is to build a generalised tangent bundle $E$ over $M$, which encodes the bosonic symmetries of supergravity. The structure group of $E$ is $\Ex{6(6)}\times\mathbb{R}^{+}$, mirroring the U-duality group of five-dimensional toroidal compactifications. The generalised tangent bundle $E$ can be written as
\begin{equation}
E \simeq TM\oplus T^{*}M\oplus T^{*}M\oplus\ext^{3}T^{*}M\oplus\ext^{5}T^{*}M \oplus\ext^{5}T^{*}M \,.
\end{equation}
The different components correspond physically to the charges of type IIB supergravity: momentum, winding, D1-, D3-, D5- and NS5-brane charge. One can combine the $T^*M$ and $\ext^{5}T^{*}M$ factors into $\SL{2;\bbR}$ doublets. This way of writing the generalised tangent bundle corresponds to the decomposition of the $\rep{27^{\prime}}$ representation of $\Ex{6(6)}$ according to a $\GL{5;\bbR}\times\SL{2;\bbR}$ subgroup, where $\GL{5;\bbR}$ acts on the five-dimensional space $M$, and $\SL{2;\bbR}$ corresponds to S-duality. A generalised vector $V$ is a section of $E$. We will use the following notation for the components of a generalised vector
\begin{equation}\label{fundGL5}
\begin{split}
\rep{27^{\prime}} & = (\rep{5},\rep{1})+(\rep{5^{\prime}},\rep{2})+(\rep{10},\rep{1})+(\rep{1},\rep{2}) \,, \\
V^M& =  \{ v^{m},\lambda_{\phantom{i}m}^{i},\rho_{mnp},\sigma_{\phantom{i}mnpqr}^{i} \} \,,
\end{split}
\end{equation}
where $m,n=1,\ldots,5$ and $i=1,2$. 

We will also need the adjoint representation of $\Ex{6(6)}\times\R^{+}$, which decomposes as
\begin{equation}\label{adjGL5}
\begin{split}
\rep{78}+\rep{1} & = (\rep{25},\rep{1})+(\rep{1},\rep{3})+(\rep{10},\rep{2})+(\rep{10^{\prime}},\rep{2})+(\rep{5},\rep{1})+(\rep{5^{\prime}},\rep{1})+(\rep{1},\rep{1}) \,, \\
\cA^{M}_{\phantom{M}N} & = \{ r_{\phantom{m}n}^{m},a_{\phantom{i}j}^{i},B_{\phantom{i}mn}^{i},\beta^{imn},C_{mnpq},\gamma^{mnpq},\ell\} \,.
\end{split}
\end{equation}
$r_{\phantom{m}n}^{m}$ is the adjoint of $\GL{5;\bbR}$, $a^{i}_{\phantom{i}j}$ is the adjoint of $\SL{2;\bbR}$, $B_{\phantom{i}mn}^{i}$ is an $\SL{2;\bbR}$ doublet of two-forms, and $C_{mnpq}$ is a four-form. In addition, there is also a doublet of bi-vectors $\beta^{imn}$ and a four-vector $\gamma^{mnpq}$. The $B^i_{\phantom{i}mn}$ and $C_{mnpq}$ fields above can be thought of as accommodating the NS-NS and R-R two-form potentials, and the R-R four-form potential of the supergravity theory.

The generalised structures $K$ and $J_\alpha$ transform under $\Ex{6(6)}\times\R^{+}$ as an element of the $\rep{27^{\prime}}$ and a triplet of elements in the $ \rep{78}$~\cite{AW15}.\footnote{We are abusing notation by not distinguishing between representations and the fibre of the corresponding vector bundle of which the objects are sections.} The $J_\alpha$ form an $\SU{2}$ triplet under the $\Ex{6(6)}$ adjoint action, corresponding to the R-symmetry of the $\cN=2$ supergravity 
\begin{equation} \label{su2algJa}
[J_\alpha,J_\beta] =2\kappa\epsilon_{\alpha\beta\gamma}J_\gamma \,,
\end{equation}
where $\kappa^2$ is the volume form on $M$. The normalisations of $K$ and $J_\alpha$ are fixed by
\begin{equation}
c(K,K,K) = \kappa^2 \,, \qqq \tr(J_{\alpha} J_\beta) =-\kappa^{2}\delta_{\alpha\beta} \,,
\end{equation}
where $c$ is the cubic invariant of $\Ex{6(6)}$, and $\tr$ is the trace in the adjoint representation (see \eqref{eq:IIB_cubic} and \eqref{eq:IIB_Killing}). The two structures are compatible, which means they satisfy
\beq
J_\alpha\cdot K=0 \,,    
\eeq
where $\cdot$ is the adjoint action on a generalised vector: $\rep{78}\times\rep{27^{\prime}}\rightarrow\rep{27^{\prime}}$ (see \eqref{eq:IIB_adjoint}).

The generalised structures $K$ and $J_\alpha$ are combinations of the geometric structures on $M$ built from bilinears of the $\mathcal{N}=2$ Killing spinors~\cite{GN16}. For Sasaki--Einstein manifolds, these are the Reeb vector $\xi$, the symplectic form $\omega$ and the holomorphic two-form $\Omega$. In this case, $K$ and $J_\alpha$ take the following simple form in terms of the $\GL{5;\bbR}\times\SL{2;\bbR}$ decompositions given in \eqref{fundGL5} and \eqref{adjGL5}~\cite{AW15}
\beq
\label{KJSE}
\begin{split}
K &= \xi -  \sigma\wedge \omega \,, \\
J_+ &= \tfrac{1}{2}\kappa u^i  (\Omega - \ii   \, \bar{\alpha} )  \,, \\
J_{3} & =\tfrac{1}{2}\kappa(  I- \ii (\sigma_{2})^{i}_{\phantom{i}j}-\tfrac{1}{4}\Omega\wedge\bar{\Omega}+\tfrac{1}{4}\bar{\alpha}\wedge\alpha ) \,,
\end{split}
\end{equation}
where $J_+=J_1 + \ii J_2$, $\sigma_{2}$ is the second Pauli matrix and the $\SL{2;\bbR}$ vector is
\beq \label{u}
u^i=(-\ii,1) \,.
\eeq
Here we are using a compact notation where we suppress all tensor indices. Thus $K$ has only vector and three-form components given by $K^m=\xi^m$ and $K_{mnp}=(\sigma\wedge\omega)_{mnp}$, whereas $J_+$ has only two-form and bi-vector components given by $(J_+)^i_{\phantom{i}mn}=\tfrac{\kappa}2 u^i \Omega_{mn}$, and  $(J_+)^{i mn}=-\ii \tfrac{\kappa}2 u^i \bar \alpha^{mn}$. Finally, $J_3$ has $\GL{5;\bbR}$ adjoint $\tfrac{\kappa}{2}I^m_{\phantom{m}n}$, $\SL{2;\bbR}$ adjoint $-\ii \tfrac{\kappa}{2} (\sigma_{2})^{i}_{\phantom{i}j}$, four-form $-\tfrac{\kappa}{8}(\Omega\wedge\bar{\Omega})_{mnpq}$ and four-vector $\tfrac{\kappa}{8}(\bar{\alpha}\wedge\alpha)^{mnpq}$ components.

Note that $K$ depends only on the Reeb vector and the contact structure, whereas $J_\alpha$ depends only on the complex structure of the K\"ahler--Einstein metric.


\subsubsection{Supersymmetry conditions}
\label{sec:susy1}

For a supersymmetric compactification to $\text{AdS}_5$, the structures $K$ and $J_\alpha$ must satisfy the differential conditions \eqref{mmconditions}--\eqref{LKK}. Let us explain a little more the form of these conditions.

The key ingredient is the generalised Lie derivative $\Dorft$. This encodes the differential geometry of the background, unifying the diffeomorphisms and gauge symmetries of the supergravity. Given two generalised vectors $V$ and $V'$ as in (\ref{fundGL5}), the generalised Lie derivative is given by  
\begin{equation}
\begin{split}\Dorft_{V}V^{\prime} & =\mathcal{L}_{v}v^{\prime}+(\mathcal{L}_{v}\lambda^{\prime i}-\imath_{v^{\prime}}\dd\lambda^{i})+(\mathcal{L}_{v}\rho^{\prime}-\imath_{v^{\prime}}\dd\rho+\epsilon_{ij}\dd\lambda^{i}\wedge\lambda^{\prime j})\\
 & \eqspace+(\mathcal{L}_{v}\sigma^{\prime i}-\imath_{v^{\prime}}\dd\sigma^{i}+\dd\rho\wedge\lambda^{\prime i}-\dd\lambda^{i}\wedge\rho^{\prime})\,,
\end{split}
\label{eq:IIB_Dorf_vector}
\end{equation}
where ${\cal L}$ is the ordinary Lie derivative. This can be extended to an action on any generalised tensor. For example, the action  on the adjoint representation is given in (\ref{eq:IIB_Dorf_adjoint}). One always has the choice to include the supergravity fluxes in the structures $K$ and $J_\alpha$ or as a modification of the generalised Lie derivative. Here the latter option turns out to be more convenient. This defines a ``twisted generalised Lie derivative" $\Dorf$, which takes the same form as (\ref{eq:IIB_Dorf_vector}) but with the substitutions 
\begin{equation}
\dd\lambda^{i}\rightarrow\dd\lambda^{i}-\imath_{v}F_{3}^{i}\,,\qqq\dd\rho\rightarrow\dd\rho-\imath_{v}F_{5}-\epsilon_{ij}\lambda^{i}\wedge F_{3}^{j}\,.\label{eq:IIB_twisted_dorf}
\end{equation}

Although we will not discuss the details, there is actually a natural hyper-K\"ahler geometry on the space of $J_\alpha$ structures \cite{AW15}. There is also an action of generalised diffeomorphisms taking one $J_\alpha$ into another. This action preserves the hyper-K\"ahler structure. The conditions \eqref{mmconditions} can then be viewed as moment maps for the action of the generalised diffeomorphisms. By construction the space $\widetilde{\cal M}$ of solutions to this condition in (\ref{tildeM}) is also hyper-K\"ahler. The generalised Lie derivative condition \eqref{genLieJa} takes a K\"ahler slice of this space. Note that for the SE structure \eqref{KJSE} and five-form flux given in \eqref{F5}, we have
\beq\label{K_Lie}
\Dorf_K = {\cal L}_\xi \, ,
\eeq
and thus $\Dorf_K$ generates the $\Uni{1}_\text{R}$ symmetry. We have shown this for the particular case of SE, but this is actually a general result. Thus the slice taken by condition (\ref{genLieJa}) essentially fixes the R-charge of $J_+$ to be $+3$, and $J_3$ to be zero.

The supersymmetry conditions can also be viewed as the internal counterpart of the supersymmetry conditions in five-dimensional gauged supergravity \cite{LST12}: \eqref{mmconditions} comes from the gravitino and gaugino variations (as does \eqref{LKK}), while \eqref{genLieJa} is related to the hyperino variation (recall $K$ is associated to the vector multiplets, while $J_\alpha$ is associated to the hypermultiplets). 

As discussed in~\cite{AW15b}, it is easy to show that the structures in \eqref{KJSE} defined for Sasaki--Einstein manifolds satisfy the supersymmetry conditions \eqref{mmconditions}--\eqref{LKK}. The first two reduce to \eqref{eq:dsigma}, \eqref{eq:dOmega} and \eqref{scalingomega}, thus fixing the constants $\lambda_\alpha$ as in \eqref{lambdaa}, while condition \eqref{LKK} gives no extra equations. Note that since the deformations we are after leave the structure $K$ invariant, the latter condition will play no role in the following.   


\subsection{Linearized deformations}

The structures $K$ and $J_\alpha$ lie in orbits of the $\Ex{6(6)}$ action. The linearized deformations are therefore elements in the adjoint of $\Ex{6(6)}$, which take us from a given point in these orbits corresponding to the original solution (in the case of Sasaki--Einstein, this is (\ref{KJSE})), to another point in the orbit corresponding to the structures of the deformed geometry.    
We have seen that from the gauge theory we expect the marginal deformations ${\cal A}$ to leave the structure $K$ invariant, while deforming $J_\alpha$.  This implies
\begin{equation}
\label{lincon}
\delta K= \cA \cdot K = 0 \,, \qqq  \delta J_\alpha= [ \cA, J_\alpha ]  \neq 0 \,.
\end{equation}
As we will discuss in more detail in appendix \ref{sec:technicalities}, the deformations $\cA$ are doublets under the $\SU2$ generated by $J_\alpha$ 
\begin{equation}
\label{cASU2}
   \cA = \begin{pmatrix} \cA^{(r)}_- \\ \cA^{(r-2)}_+ \end{pmatrix} \, ,
\end{equation}
with  $\mathcal{A}_-  = [ J_+ , \mathcal{A}_+ ]$.\footnote{Strictly speaking, this should be $\mathcal{A}_-  =\kappa^{-1} [ J_+ , \mathcal{A}_+ ]$, but we have dropped the factors of $\kappa$ for ease of presentation in this section.} The signs  $\pm$ denote the charge under $J_3$, $[J_3, \mathcal{A}_\pm ] = \pm \ii \mathcal{A}_\pm$, and $r$ is the charge under the action of $\Dorf_K$ corresponding to their R-charge
\begin{equation}
\label{ARcharge}
   \Dorf_K \cA^{(r)}_\pm = \tfrac32\ii r\cA^{(r)}_\pm \, .
\end{equation}
The difference in the R-charge of the two components follows from \eqref{genLieJa}, \eqref{ARcharge} and the definition $\mathcal{A}_-  = [ J_+ , \mathcal{A}_+ ]$.

We now need to find pairs of solutions for $\cA_\pm$ satisfying the linearised supersymmetry conditions and, for definiteness, R-charge $r\geq0$. In the next subsection, we start by first finding solutions to 
the linearised moment  maps.  We then have to mod out by the symmetry, identifying deformations that are related by diffeomorphisms or form-field gauge transformations as corresponding to the same physical deformation. This process corresponds to finding the bulk modes dual to the bosonic components of all chiral superfields: namely the chiral ring operators $A_i$ (associated to $\mathcal{A}_-$) and the related supersymmetric deformations of the Lagrangian $F_{A_i}$ (associated to $\mathcal{A}_+$). Then in the following subsection, we turn to finding the subset of marginal deformations. The technical details are discussed in appendix~\ref{sec:technicalities}. Here we outline the procedure and present the results. 

\subsubsection{The chiral ring}
\label{sec:chiral-ring}

The linearised moment map equations are given by\footnote{As we discuss in appendix \ref{susy_section}, the actual deformation is by $\mathcal{A}=\re \mathcal{A}_+$ so that the deformed structures are real. This do not affect the discussion that follows.} 
\begin{align}
\label{deltamuaB}
\delta \mu_\alpha(V)  = \int \kappa \tr  (J_\alpha, \Dorf_V \cA ) =0 \qquad \forall \, V \in \rep{27^\prime} \,,
\end{align}
where we are using the fact that the deformation leaves $K$ invariant. 

We start by looking for $\cA_+$ that solve \eqref{deltamuaB}.  The $\cA_+$  deformations  can be distinguished by which components of the $\Ex{6(6)}\times\mathbb{R}^{+}$ adjoint~\eqref{adjGL5} are non-zero. They fall into two classes
\beq
\label{AAtform}
\Ac_{+}  = B^{i} + \beta^{i} \,, \qqq 
\Ah_{+}  =  a^i_{\phantom{i}j} \, ,
\eeq
where the first contains only two-forms and the corresponding bi-vectors, and the second contains only $\sln{2}$ entries.

As shown in appendix \ref{sec:mmtech}, the two-form part of the $\Ac_{+}$ solutions to \eqref{deltamuaB} consists of two independent terms
\begin{equation}
   B^{i} = - \tfrac12 \ii\bar u^i \left[  f \bar \Omega  + \tfrac{1}{2q(q-1)}\partial(\partial f \lrcorner\bar{\Omega}) +  \tfrac{\ii}{q} \sigma \wedge (\partial f \lrcorner\bar{\Omega})    \right]  - \ii\bar u^i \delta    \,  ,
 \end{equation}
where  $\Omega$ and $\sigma$ are the holomorphic two-form and the contact form on the SE manifold, and the $\SL{2;\bbR}$ vector $u^i$ is defined in (\ref{u}). The bi-vector part of the solution is obtained by raising indices with the SE metric. The term in the brackets is completely determined by a function $f$  on the SE manifold satisfying 
\beq
\label{fcond}
\bar{\partial} f = 0\, , \qqq \mathcal{L}_\xi f = \ii q f\, .
\eeq
Note that $f$ is holomorphic with respect to $\bar{\partial}$ if and only if it is the restriction of a holomorphic function on the Calabi--Yau cone over the Sasaki--Einstein base~\cite{EST13}. The second term depends only on a primitive $(1,1)$-form $\delta$ on the KE base that is closed under both $\partial$ and $\bar{\partial}$
\beq
\label{deltacond}
 \delta\wedge\omega=0 \, , \qqq \del\delta=\bar{\del}\delta = 0 \, .
\eeq
Imposing that the deformation $ \Ac_{+} $ has fixed R-charge $r-2$, and using \eqref{scalingomega}, gives
\begin{equation}
\label{r1}
   \mathcal{L}_\xi f = \tfrac32\ii r f \, , \qqq
   \mathcal{L}_\xi \delta =  \tfrac32 \ii (r-2) \delta  \,  ,
\end{equation}
so that $f$ is a homogeneous function on the Calabi--Yau cone of degree $\frac32 r$.

Let us now consider $ \Ah_{+}$. Its only non-zero components are $a^i_{\phantom{i}j} \in \sln{2}$, which are again determined by a function $\tilde f$ on the manifold
\beq
\label{ft}
\Ah_+ = -\tfrac{1}{2} \tilde{f} \bar{u}^i \bar{u}_j \,,
\eeq
where $\bar u_i = \epsilon_{ij}\bar u^j$ and the function $\tilde{f}$ is holomorphic
\beq
\label{ftcond}
\bar{\partial} \tilde f = 0 \, . 
\eeq
The deformations of fixed R-charge $r-2$ satisfy 
\begin{equation}
\label{r2}
   \mathcal{L}_\xi \tilde f  = \tfrac32\ii (r-2) \tilde f \, ,
\end{equation}
so that $\ft$ is a homogeneous function on the Calabi--Yau cone of degree $\frac32 (r-2)$.

For each solution $\cA_+$, one can generate an independent solution $\cA_-$ by acting with $J_+$. Indeed, any deformation of the form $\cA_-=[J_+,\cA_+]$ is automatically a solution of the moment maps, provided $\cA_+$ is. The explicit form of these deformations for $\Ac_-$ and $\Ah_-$ is given in \eqref{Acrminus} and \eqref{Ahatminus}. Thus the solutions of the linearised  moment maps consist of an infinite set of deformations  $\cA_+$  labelled by their R-charge $r$, which are generated by the two holomorphic functions, $f$ and $\tilde f$, and a (1,1)-form, $\delta$,
and another independent set of deformations $\cA_-$  generated by $f'$, $\tilde f'$ and $\delta'$. Together these give the general solution to the deformation problem.  Arranging these deformations  as in \eqref{cASU2}, we find three types of multiplets, schematically, 
\beq
\label{defmult}
  \begin{pmatrix} \cA^{(r)}_- \\ \cA^{(r-2)}_+ \end{pmatrix}   \sim   \begin{pmatrix} f' \\ f  \end{pmatrix} \, , 
   \,\, \begin{pmatrix} \tilde f^\prime  \\ \tilde f \end{pmatrix}  \, , 
    \,\,\begin{pmatrix} \delta' \\ \delta  \end{pmatrix} \, ,
 \eeq     
with charge $r$ given respectively by $r >0$, $r \geq 2$ and $r=2$. 

Let us now identify what these solutions correspond to physically. For this it is convenient to compute the action of the linearised deformations on the bosonic fields of type II supergravity and then interpret the multiplets \eqref{defmult} in terms of Kaluza--Klein modes on the Sasaki--Einstein manifold. One way to read off the bosonic background is from the generalised metric $G$. This is defined in \eqref{gen_metric} and encodes the metric, dilaton, the NS-NS field $B_2$ and the R-R fields $C_0$, $C_2$ and $C_4$. As discussed in appendix \ref{sec:genmet}, the two-form and bi-vector deformations $f$ and their partners $f^\prime$ at leading order generate NS-NS and R-R two-form potentials, and a combination of internal four-form potential and metric\footnote{The full form of the four-form potential and metric is given by \eqref{Acrminus} with $\bar{\nu}^\prime = \tfrac{\ii}{2q} \partial f^\prime \lrcorner \bar{\Omega}$ and $\hat{\omega}^\prime = \tfrac{1}{4q(q-1)} \partial(\partial f^\prime \lrcorner \bar{\Omega})$.}
\beq
\label{mult1}
 \begin{pmatrix} f' \\ f  \end{pmatrix} \sim \begin{pmatrix}
 C_4  + g^{a}_{\phantom{a}a} \\ 
 C_{2}-\ii  B_{2} 
 \end{pmatrix}
 \propto
    \begin{pmatrix} \frac{1}{2} f^\prime \Omega \wedge \bar{\Omega} +\tfrac{\ii}{2q}\Omega\wedge\sigma\wedge(\partial f^\prime \lrcorner \bar{\Omega}) +\ldots \\   f \bar \Omega  + \tfrac{1}{2q(q-1)}\partial(\partial f \lrcorner\bar{\Omega}) +  \tfrac{\ii}{q} \sigma \wedge (\partial f \lrcorner\bar{\Omega}) \end{pmatrix}   .
\eeq
Similarly one can show that the holomorphic function $\tilde f$ and its partner $\tilde f^\prime$ correspond to the axion-dilaton, and NS-NS and R-R two-form potentials
\beq
\label{mult2}
 \begin{pmatrix} \tilde f^\prime  \\ \tilde f \end{pmatrix}   \sim   
 \begin{pmatrix} C_{2}-\ii  B_{2}   \\  C_0 - \ii \phi   \end{pmatrix}
 \propto
 \begin{pmatrix} \tilde{f}^\prime \Omega   \\  \tilde f  \end{pmatrix}   . 
\eeq
Finally the two-form and bi-vector deformations  $\delta$  and its partner $\delta^\prime$ generate NS-NS and R-R two-form potentials and a component of the internal metric
\beq
\label{mult3}
 \begin{pmatrix} \delta'  \\  \delta   \end{pmatrix} \sim \begin{pmatrix}
 g^m_{\phantom{m}n} \\ 
 C_{2}-\ii  B_{2} 
 \end{pmatrix}
 \propto
  \begin{pmatrix}   (j\bar{\alpha} \lrcorner j \delta^\prime + j \delta^\prime \lrcorner j \Omega)^m_{\phantom{m}n}  \\   \delta  \end{pmatrix}   .
\eeq

The KK spectrum for a generic Sasaki--Einstein background was analysed in~\cite{EST13} by solving for eigenmodes of the Laplacian on the manifold. The states arrange into long and short multiplets of $\mathcal{N}=2$ supergravity in five dimensions. Our multiplets \eqref{mult1}, \eqref{mult2} and \eqref{mult3} are indeed the short multiplets of \cite{EST13}.

In terms of the bulk five-dimensional supergravity, each $(\cA^{(r)}_-,\cA^{(r-2)}_+)$ doublet of fixed R-charge corresponds to a different hypermultiplet. In the dual field theory the $\cA^{(r-2)}_+$ piece corresponds to the $\theta^2$-component of a chiral superfield while the $\cA^{(r)}_-$ piece corresponds to the lowest component~\cite{Tachikawa06}. We then have the following mapping between supergravity and field theory multiplets 
\begin{equation}
\label{ringsol}
 \begin{array}{ll}
 \begin{pmatrix} f' \\ f \end{pmatrix} 
       \sim \tr \mathcal{O}_f \,, &    \text{superpotential deformations, $r>0$,} \\[7pt] \\ 
 \begin{pmatrix}  \tilde f' \\ \tilde f  \end{pmatrix}
        \sim \tr W_\alpha W^\alpha \mathcal{O}_{\tilde f} \,, \,\,\,\,\,
        & \text{coupling deformations, $r\geq 2,$} \\[7pt] \\
 \begin{pmatrix}  \delta' \\ \delta 
           \end{pmatrix}
       \sim \mathcal{O}_{\text{gauge}} \,,& 
       \text{difference in gauge couplings, $r=2$.}
 \end{array}
\end{equation}
For $\text{S}^5$ the first two sets of multiplets corresponds to the operators $\tr (\Phi^k)$ and $\tr (  W_\alpha W^\alpha \Phi^k)$, where $\Phi$ denotes any of the three adjoint chiral superfields of $\mathcal{N}=4$ SYM, and the last multiplet is not present. For $\text{T}^{1,1}$, one has $\tr (\mathcal{O}_f) =  \tr(AB)^k $,  $\tr (W_\alpha W^\alpha \mathcal{O}_{\tilde f})  =  \tr\bigl[(W^2_A +  W^2_B) (AB)^k\bigr]$ and $\mathcal{O}_{\text{gauge}} =  \tr (W^2_A -  W^2_B)$ where $A$ and $B$ denote the two doublets of bi-fundamental chiral superfields. 
In analogy with the $\text{T}^{1,1}$ case, for a generic SE the operators $\mathcal{O}_f$ and $\mathcal{O}_{\tilde f}$ are products of chiral bi-fundamental superfields of the theory, while $\mathcal{O}_{\text{gauge}}$ corresponds to changing the relative couplings of the gauge groups. 

The tower of deformations gives the space $\widetilde \cM$ defined in \eqref{Mtilde}. In particular, the $\cA_- =  ( f', \tilde f', \delta') \sim A_i $ deformations parametrise the chiral ring, while $\cA_+ =  (f, \ft, \delta)  \sim F_{A_i}$ parametrise the superpotential deformations.

\subsubsection{Marginal deformations}
\label{sec:marg-deform}

The marginal deformations are a subspace of solutions in $\widetilde{\mathcal M}$ that also satisfy the second differential condition~\eqref{genLieJa}. At first order in the deformation, this is
\beq \label{genLieJafirst}
 [ \Dorf_K \cA, J_\alpha ] = 0  \, ,
\eeq
where we have used again the fact that the deformations leave $K$ invariant. Since the commutators with $J_\alpha$ are non-zero, this condition amounts to the requirement 
\begin{equation}
\label{gen-degrees}
   \Dorf_K \mathcal{A} = 0 \qquad \Rightarrow \qquad  
   \mathcal{L}_\xi \mathcal{A} = 0 \, . 
\end{equation}
In other words, the R-charge of $\mathcal{A}$ vanishes. Comparing with~\eqref{defmult} we see that the $\cA_-$ components always have positive R-charge and therefore
are not solutions of  \eqref{genLieJafirst}. Thus marginal deformations can only be given by  the $\cA^{(r-2)}_+$ components with  $r=2$. This is consistent, because, as we have mentioned, the $\cA_+$ components correspond to deforming the SCFT by $\theta^2$ terms, which are supersymmetric, whereas the $\cA_-$ terms correspond to the lowest component of a chiral superfield and so do not give supersymmetric deformations. 

From~\eqref{r1} and~\eqref{r2} we see that the $\cA_+^{(0)}$ components ($r=2$) are\footnote{$\text{H}_{\text{prim}}^{1,1}(M)$ denotes the cohomology of primitive $(1,1)$-forms.}
\begin{equation} \label{degrees}
   \text{$f$ of degree 3} \, , \qqq
   \tilde f  = \text{constant} \, , \qqq
   \delta\in \text{H}_{\text{prim}}^{1,1}(M) \, ,
\end{equation}
corresponding precisely to superpotential deformations with $\Delta=3$, a change in  the original superpotential (and at the same time of the sum of coupling constants), and a change in  the relative gauge couplings respectively.

\subsubsection{Linearised supergravity solution}

We want now to compute the supergravity solutions at linear order. As discussed in detail in appendix  \ref{sec:genmet}, this can be done by looking at the action of the marginal deformations $\Ac_+$ and $\Ah_+$ on the generalised metric, which encodes the bosonic fields  of type IIB supergravity. We first consider the effect of a $\Ac_+$ deformation to linear order. As already mentioned, such a deformation generates NS-NS and R-R two-form potentials, given by
\begin{equation}
 C_{2}-\ii  B_{2} = -\ii \bigl(f \bar{\Omega} + \tfrac{1}{12}\partial(\partial f\lrcorner\bar{\Omega}) + \tfrac{1}{3} \ii \sigma \wedge (\partial f\lrcorner\bar{\Omega}) \bigr) - 2 i \delta \, .
\end{equation}
Taking an exterior derivative, the complexified flux $G_{3}=\dd(C_{2}-\ii B_{2})$ to leading order is
\begin{equation} \label{G3}
G_{3} = - \tfrac{4}{3}\ii \partial f \wedge \bar{\Omega} + 4 f \sigma \wedge \bar{\Omega} - \tfrac{1}{3} \sigma \wedge \partial (\partial f\lrcorner\bar{\Omega}) \, .
\end{equation}
The (1,1)-form $\delta\in \text{H}^2(M)$ is closed and therefore does not contribute to the flux. On the Calabi--Yau cone, it is well-known that superpotential deformations correspond to imaginary anti-self-dual (IASD) flux~\cite{GP01}. The $G_3$ here is the component of the IASD flux restricted to the Sasaki--Einstein space.

Now consider the effect of a marginal $\Ah_+$ deformation to linear order. As we show in appendix \ref{axiondilaton}, such a deformation allows for non-zero, constant values of the axion and dilaton, given by
\begin{equation}\label{tildefaxiondilaton}
\tilde{f}=C_0 - \ii \phi \, .
\end{equation}

We stress that this calculation and the expressions for the leading-order corrections to the solution~\eqref{G3} for the NS-NS and R-R three-form flux and the axion-dilaton in~\eqref{tildefaxiondilaton} are valid for \emph{any} Sasaki--Einstein background. One simply needs to plug in the expressions for the holomorphic form and contact structure of the given Sasaki--Einstein space. These objects are given in terms of a frame in \eqref{frame}. We will give the explicit form of the frame for the examples of $\text{S}^{5}$, $\text{T}^{1,1}$ and the $\text{Y}^{p,q}$ manifolds, and compare the flux with some known results in section \ref{examples}.


\subsection{Moment maps, fixed points and obstructions}
\label{sec:obstructions}

The linearised analysis above has identified the supergravity perturbations dual to marginal chiral operators in the SCFT. However, this is not the end of the story. Really we would like to find the exactly marginal operators. In the gravity dual this means solving the supersymmetry equations not just to first order but to all orders. In general there are obstructions to solving the supersymmetry conditions to higher orders, and not all marginal deformations are exactly marginal~\cite{AKY02}. As we saw in section~\ref{sec:scft}, in the field theory these obstructions are related to global symmetries~\cite{GKSTW10}.

As we discussed in section~\ref{sec:exact}, the fact that the supergravity conditions in exceptional generalised geometry appear as moment maps gives an elegant interpretation of the field theory result. This analysis was completely generic,  equally applicable to type II and eleven-dimensional supergravity backgrounds. We will now give a few more details, using the Sasaki--Einstein case as a particular example. 

The key point is that generically there are no obstructions to extending the linearised solution of a moment map to an all-orders solution. The only case when this fails is when one is expanding around a point where some of the symmetries defining the moment map have fixed points (see for instance~\cite{Thomas06}). Since here the moment maps are for the generalised diffeomorphisms, we see that there are obstructions only when the background is invariant under some subgroup $G$ of diffeomorphisms and gauge transformations, called the stabiliser group. Such transformations correspond to additional global symmetries in the SCFT. Furthermore, one can use a linear analysis around the fixed point to show that the obstruction appears as a further symplectic quotient by the symmetry group $G$. This mirrors the field theory result that all marginal deformations are exactly marginal unless there is an enhanced global symmetry group and that the space of exactly marginal operators is a symplectic quotient of the space of marginal operators. 

To see this in a little more detail let us start by reviewing how the conditions~\eqref{mmconditions} appear as moment maps~\cite{AW15,AW15b} and how the obstruction appears. We will first consider $\widetilde{\mathcal{M}}$, the space of chiral ring elements and $\theta^2$-components, and then at the end turn to the actual marginal deformations. As we stressed above, this discussion is completely generic and not restricted to Sasaki--Einstein spaces. One first considers the space $\MH^K$ of all possible hypermultiplet structures compatible with a fixed $K$, in other words
\begin{equation}
   \MH^K = \{ J_a(x) : J_a\cdot K = 0 \} .
\end{equation}
Since each point $p\in\MH^K$ is a choice of structure defined by a triplet of functions $J_\alpha(x)$ on $M$, the space $\MH^K$ is infinite dimensional. Nonetheless it is hyper-K\"ahler. A tangent vector $v$ at the point $p$ can be thought of as a small change in the structure 
\begin{equation}
   v_\alpha(x) = \delta J_\alpha(x) = [ \mathcal{A}(x) , J_\alpha(x) ] 
      \in T_p\MH^K \, ,
\end{equation}
where $\mathcal{A}(x)$ is some $\Ex{6(6)}\times\bbR^+$ element. The hyper-K\"ahler structure is characterised by a triplet of closed symplectic forms, $\Omega_\alpha$. These symplectic structures $\Omega_\alpha$ are defined such that, given a pair of tangent vectors $v,v'\in T_p\MH^K$, the three symplectic products are given by
\begin{equation}
\label{HK-X}
   \Omega_\alpha(v,v') = \epsilon_{\alpha\beta\gamma} \int \tr (v_\beta v'_\gamma)
       = 2 \int \kappa\tr \bigl([\mathcal{A},\mathcal{A}'] J_\alpha \bigr)\, . 
\end{equation}

The generalised diffeomorphism group acts on $J_\alpha(x)$ and hence on $\MH^K$. Furthermore its action leaves the symplectic forms $\Omega_\alpha$ invariant. Infinitesimally, generalised diffeomorphisms are generated by the generalised Lie derivative so that $\delta J_\alpha=\Dorf_VJ_\alpha\in T_p X$. Thus, just as vector fields parametrise the Lie algebra of conventional diffeomorphisms via the Lie derivative, one can view the generalised vectors $V$ as parametrising the Lie algebra $\gdiff$ of the generalised diffeomorphism group.\footnote{Note from~\eqref{eq:IIB_Dorf_vector} that shifting the form components $\lambda^i$ and $\rho$ of $V$ by exact terms does not change $\Dorf_V$, furthermore it is independent of $\sigma^i$. Thus different generalised vectors can parametrise the same Lie algebra element.} One can then show that the $\mu_\alpha(V)$ in~\eqref{mmconditions} are precisely the moment maps for the action of the generalised diffeomorphism group on $\MH^K$. As written they are three functions on $\MH^K\times\gdiff$ where $J_\alpha$ gives the point in $\MH^K$ and $V$ parametrises the element of $\gdiff$, but they can equally well be viewed as a single map $\mu\colon \MH^K\to\gdiff^*\times\bbR^3$ where $\gdiff^*$ is the dual of the Lie algebra. Solving the moment map conditions~\eqref{mmconditions} and modding out by the generalised diffeomorphisms to obtain $\widetilde{\mathcal{M}}$ as in~\eqref{tildeM} is a hyper-K\"ahler quotient. As discussed in~\cite{AW15}, one subtlety is that, in order to define a quotient, the right-hand side of the conditions $\lambda_\alpha\gamma$, given in~\eqref{alphadef} and which depends on $K$, must be invariant under the action of the group. Thus the quotient is really defined not for the full generalised diffeomorphism group, but rather the subgroup $\GDiff_K$ that leaves $K$ invariant. Infinitesimally $V$ parametrises an element of the corresponding Lie algebra $\gdiff_K$ if $\Dorf_VK=0$. Thus we have the quotient~\eqref{tildeM}.

The linearised analysis of the last section first fixes a point $p\in \MH^K$ corresponding to the Sasaki--Einstein background satisfying the moment map conditions, and then finds deformations of the structure $\delta J_\alpha\in T_p\MH^K$ for which the variations of the moment maps $\delta \mu_\alpha(V)$ vanish for all $V$. If we view $\delta\mu_\alpha$ as a single map $\delta\mu\colon T_p\MH^K\to \gdiff_K^*\times\bbR^3$, the linearised solutions live in the kernel. Suppose now that $p$ is fixed under some subset of generalised diffeomorphisms, that is we have a stabiliser group $G\subset\GDiff_K$. The corresponding Lie subalgebra $\mathfrak{g}\subset\gdiff_K$ is 
\begin{equation}
   \mathfrak{g} = \{ V\in \gdiff_K: \Dorf_V J_\alpha =0 \} \, . 
\end{equation}
At a generic point in $\MH^K$ satisfying the moment map conditions, all elements of $\GDiff_K$ act non-trivially and so the stabiliser group is trivial. Thus solving $\delta\mu_\alpha(V)=0$ we get a constraint for every $V\in\gdiff_K$. In constrast, at the point $p$, we miss those constraints corresponding to $V\in\mathfrak{g}$. Thus we see that the obstruction to extending the first-order deformation to all orders lies precisely in $\mathfrak{g}^*\times\bbR^3$, that is, it is the missing constraints. Put more formally,\footnote{See for example the note in section 5 of~\cite{Thomas06}.} the embedding $i\colon\mathfrak{g}\to\gdiff_K$ induces a map $i^*\colon\gdiff_K^*\to\mathfrak{g}^*$ on the dual spaces and, at $p$, we have an exact sequence   
\begin{equation}
   \begin{tikzcd}
       T_p\MH^K \arrow{r}{\delta\mu}
         & \gdiff_K^{*}\times\bbR^3 \arrow{r}{i^*}
         & \mathfrak{g}\times\bbR^3  
   \end{tikzcd} . 
\end{equation}
The map $\delta\mu$ is not onto and the obstruction is its cokernel $\mathfrak{g}^*\times\bbR^3$. 

The standard argument for moment maps at fixed points actually goes further. Let $\mathcal{U}$ be the vector space of linearised solutions $\delta\mu_\alpha(V)=0$ at $p$, up to gauge equivalence. For the Sasaki--Einstein case it is the space of solutions, dual to the couplings of the operators $(A_i,F_{A_i})$, given in~\eqref{ringsol}. Formally $\mathcal{U}$ is defined as follows. Recall that the space of solutions is $\ker\delta\mu\subset T_p\MH^K$. The action of $\GDiff_K$ on $p\in \MH^K$ defines an orbit $O\subset \MH^K$, and modding out by the tangent space to the orbit $T_p O$ at $p$ corresponds to removing gauge equivalence, so that
\begin{equation}
   \mathcal{U} = \ker\delta\mu \quotient T_pO \, . 
\end{equation}
The moment map construction means that the hyper-K\"ahler structure on $T_p\MH^K$ descends to $\mathcal{U}$. By definition, the stabiliser group $G$ acts linearly on $T_p\MH^K$ and this also descends to $\mathcal{U}$. Furthermore it preserves the hyper-K\"ahler structure. Thus we can actually define moment maps $\tilde{\mu}_\alpha$ for the action of $G$ on $\mathcal{U}$. The standard argument is then that the space of unobstructed linear solutions can be identified with the hyper-K\"ahler quotient of $\mathcal{U}$ by $G$, so near $p$ we have 
\begin{equation}
   \widetilde{\mathcal{M}}
      = \mathcal{U} \qqquotient G 
      \coloneqq \frac{\{\mathcal{A}\in \mathcal{U}: 
          \tilde{\mu}_\alpha=0\}}{G} \, , 
\end{equation}
just as in~\eqref{Mtilde}. The idea here is that if we move slightly away from $p$ we are no longer at a fixed point and there are no missing constraints. Thus we really want to take the hyper-K\"ahler quotient in a small neighbourhood of $\MH^K$ near $p$. However we can use the tangent space $T_p\MH^K$ to approximate the neighbourhood. The moment map on $T_p\MH^K$ can be thought of in two steps: first we impose $\delta\mu_\alpha=0$ at the origin and mod out by the corresponding gauge symmetries, reducing $T_p\MH^K$ to the space $\mathcal{U}$. However this misses the conditions coming from the stabiliser group $G$ which leaves the origin invariant. Imposing these conditions takes a further hyper-K\"ahler quotient of $\mathcal{U}$ by $G$. Finally, note that since $G$ acts linearly on $\mathcal{U}$, the obstruction moment maps $\tilde{\mu}_\alpha$ are quadratic in the deformation $\mathcal{A}$. This exactly matches the analysis in~\cite{AKY02}, where in solving the deformation to third-order the authors found a quadratic obstruction. What is striking is that we have been able to show how the obstructions appear for completely generic supersymmetric backgrounds. 

This discussion has been somewhat abstract. Let us now focus on the simple case of $\text{S}^5$ to see how it works concretely. The full isometry group is $\SO6\simeq\SU4$. However, only an $\SU3$ subgroup preserves $J_\alpha$ and $K$, hence 
\begin{equation*}
   \text{for $\text{S}^5$ the stabiliser group is $G=\SU3$.}
\end{equation*}
Rather than consider the full space of linearised solutions~\eqref{ringsol}, for simplicity we will just focus on $f$ and $f^\prime$, and furthermore assume both functions are degree three:  $\mathcal{L}_\xi f=3\ii f$ and $\mathcal{L}_\xi f^\prime=3\ii f^\prime$. In terms of holomorphic functions on the cone $\bbC^3$, this implies both functions are cubic
\begin{equation}
   f = f^{ijk} z_iz_jz_k \, , \qqq
   f' = f^{\prime ijk} z_iz_jz_k \, .
\end{equation}
The coefficients $(f^{ijk},f^{\prime ijk})$ parametrise a subspace in the space of linearised gauge-fixed solutions $\mathcal{U}$. Using the expressions~\eqref{AAtform} and~\eqref{HK-X} one can calculate the hyper-K\"ahler metric on the $(f^{ijk},f^{\prime ijk})$ subspace. Alternatively, one notes that the hyper-K\"ahler structure on $\MH^K$ descends to a flat hyper-K\"ahler structure the subspace, parametrised by $f^{ijk}$ and $f^{\prime ijk}$ as quaternionic coordinates. We then find the three symplectic forms
\begin{equation}
\begin{split} 
   \Omega_3 
      &= \tfrac12\ii \,\dd f^{ijk} \wedge \dd \bar{f}_{ijk} 
         - \tfrac12\ii \,\dd f^{\prime ijk} 
            \wedge \dd \bar{f}^{\prime}_{ijk} \, , \\
   \Omega_+ &= \dd f^{ijk} \wedge \dd \bar{f}^{\prime}_{ijk} \, , 
\end{split}
\end{equation}
where $\Omega_+=\Omega_1+\ii\Omega_2$ and indices are raised and lowered using $\delta_{ij}$. The $\SU3$ group acts infinitesimally as 
\begin{equation}
\begin{split}
   \delta f^{ijk} &= a^{[i}{}_l f^{jk]l} \, , \\
   \delta f^{\prime ijk} &= a^{[i}{}_l f^{\prime jk]l} \, ,
\end{split}
\end{equation}
where $\tr a=0$ and $a^\dag=-a$. This action is generated by the vectors
\begin{equation}
   \rho(a) = a^i{}_j \big( 
      f^{jkl}\del_{ikl} - \bar{f}_{ikl}\bar{\del}^{jkl}
      + f^{\prime jkl}\del^{\prime}_{ikl} - \bar{f}^{\prime}_{ikl}\bar{\del}^{\prime jkl}
      \big) \, , 
\end{equation}
where $\del_{ijk}=\del/\del f^{ijk}$ and $\del^{\prime}_{ijk}=\del/\del f^{\prime ijk}$. It is then easy to solve for the (equivariant) moment maps $\tilde{\mu}_\alpha(a)$ satisfying $i_{\rho(a)}\Omega_\alpha=\dd\tilde{\mu}_\alpha(a)$, to find
\begin{equation}
\begin{split}
   \tilde{\mu}_3(a) &= \tfrac12 \ii a^i{}_j \big( 
      f^{jkl}\bar{f}_{ikl} - f^{\prime jkl}\bar{f}^{\prime}_{ikl} \big) \, , \\
   \tilde{\mu}_+(a) &= a^i{}_j f^{jkl}\bar{f}^{\prime}_{ikl} \, . 
\end{split}   
\end{equation}
Solving the moment maps $\tilde{\mu}_\alpha(a)=0$ for all $a^i_{\phantom{i}j}$ gives
\begin{equation}
\label{tmm}
\begin{split}
   \tfrac12 \ii \big( 
      f^{ikl}\bar{f}_{jkl} 
        - \tfrac13\delta^i{}_j f^{klm}\bar{f}_{klm}
      - f^{\prime ikl}\bar{f}^{\prime}_{jkl} 
        + \tfrac13\delta^i{}_j f^{\prime klm}\bar{f}^{\prime}_{klm}
        \big) &= 0\, , \\
   f^{ikl}\bar{f}^{\prime}_{jkl} 
        - \tfrac13 \delta^i{}_j f^{klm}\bar{f}^{\prime}_{klm} 
        &= 0 \, . 
\end{split}   
\end{equation}
Imposing these conditions and modding out by $\SU3$ then gives the unobstructed deformations living in $\widetilde{\mathcal{M}}$. If we actually included all the modes in $(f,f')$ we would find polynomials with arbitrary coefficients $f^{i_1\dots i_p}$ but the construction would be essentially the same. This also applies to the $(\ft,\ft^\prime)$ modes. Since $\text{H}^2(\text{S}^5)=0$ there are no $(\delta,\delta^\prime)$ solutions on $\text{S}^5$. 

So far we have discussed how the existence of fixed points leads to obstructions in the construction of the space $\widetilde{\mathcal{M}}$. However ultimately we would like to find the unobstructed exactly marginal deformations $\mathcal{M}_\text{c}$. Returning to the generic case, recall that the marginal deformations corresponded to a subspace given by the $\mathcal{A}_+^{(0)}$ components of the full set of deformations, satisfying the condition~\eqref{gen-degrees}. (In the Sasaki--Einstein case these are given in~\eqref{degrees}.) Let us denote this subspace by $\mathcal{U}_\text{c}\subset\mathcal{U}$. Since $\Dorf_KJ_\alpha$ is a holomorphic vector on $\widetilde{\mathcal{M}}$ with respect to one of the complex structures~\cite{AW15b}, $\mathcal{U}_\text{c}$ is a K\"ahler subspace. Furthermore, taking the hyper-K\"ahler quotient by $G$ and then restricting to the marginal deformations is the same as restricting to the marginal deformations and then taking a symplectic quotient by $G$ using only the moment map $\lambda^\alpha\tilde{\mu}_\alpha$. In other words the diagram
\begin{equation}
   \begin{tikzcd}[row sep=large]
       \mathcal{U} \arrow{r} \arrow{d}
         & \mathcal{U}_{\text{c}} \arrow{d} \\
       \widetilde{\mathcal{M}}=\mathcal{U}_\text{c} \qqquotient G \arrow{r}
         & {\mathcal{M}}_\text{c}=\mathcal{U}_\text{c} \qquotient G 
   \end{tikzcd}
\end{equation}
commutes. This is because the action of $\Dorf_K$ which enters the generalised Lie derivative condition~\eqref{genLieJa} commutes with the action of $\Dorf_V$ generating $G$.\footnote{We have $[\Dorf_V,\Dorf_K]=\Dorf_{\Dorf_VK}=0$ since by definition $\Dorf_VK=0$ if $V$ is in the stabiliser group $G$.}. Given $\mathcal{U}_\text{c} \qquotient G=\mathcal{U}_\text{c} \quotient G^\bbC$, we see that we reproduce the field theory result~\eqref{Mtilde}. 

It is simple to see how this works in the case of $\text{S}^5$. The marginal modes correspond to $f^\prime=\ft^\prime=0$, while $f$ is restricted to be degree three and $\ft$ constant (recall $\delta$ and $\delta^\prime$ are absent on $\text{S}^5$). Since constant $\ft$ is invariant under $\SU3$, the moment map conditions $\tilde{\mu}_\alpha=0$ on the marginal modes reduce to a single condition that comes from $\tilde{\mu}_3$ (given $\lambda_1=\lambda_2=0$), namely 
\begin{equation}
   \tfrac12 \ii \bigl( 
      f^{ikl}\bar{f}_{jkl} 
        - \tfrac13\delta^i_{\phantom{i}j} f^{klm}\bar{f}_{klm} \bigr) = 0 \, , 
\end{equation}
since the $\tilde{\mu}_+$ moment map is satisfied identically as $f^\prime=\ft^\prime=0$. Comparing with section~\ref{sec:scft}, we see that we indeed reproduce the field theory result that the exactly marginal deformations are a symplectic quotient of the marginal deformations by the global symmetry group $G$.


\section{Examples} \label{examples}

In the previous section we derived the first-order supergravity solution dual to exactly marginal deformations on any Sasaki--Einstein background. We now apply this to the explicit examples of the supergravity backgrounds dual to $\cN=4$ super Yang--Mills, the $\cN=1$ Klebanov--Witten theory and $\cN=1$ $\text{Y}^{p,q}$ gauge theories.

\subsection{\texorpdfstring{$\cN=4$}{N=4} super Yang--Mills}

The Sasaki--Einstein manifold that appears in the dual to $\cN =4$ SYM is $\text{S}^5$, whose four-dimensional K\"ahler--Einstein base is $\mathbb{CP}^2$. The metric on $\text{S}^{5}$ can be written as\footnote{Here $s_{\alpha}$ and $c_\alpha$ are shorthand for $\sin \alpha$ and $\cos \alpha$, and similarly for $\theta$.} 
\begin{equation}\label{S5_metric}
\dd s^{2}(\text{S}^{5})=\dd\alpha^{2}+s_{\alpha}^{2}\dd\theta^{2}+c_{\alpha}^{2}\dd\phi_{1}^{2}+s_{\alpha}^{2}c_{\theta}^{2}\dd\phi^2_{2}+s_{\alpha}^{2}s_{\theta}^{2}\dd\phi^2_{3} \,,
\end{equation}
where the coordinates are related to the usual complex coordinates on $\bbC^3$, pulled back to $\text{S}^{5}$, by
\begin{equation}
\label{compl}
 z_{1}  = c_\alpha \ee^{\ii\phi_1} \,, \qquad
 z_{2}  = s_\alpha c_\theta \ee^{\ii\phi_2} \,, \qquad
 z_{3} = s_\alpha s_\theta \ee^{\ii\phi_3} \, .
\end{equation}

We can take the following frame for $\text{S}^5$
\begin{equation}\label{frameS5}
\begin{split}
e_{1} & =c_{\alpha}^{2}\dd\phi_{1}+c_{\theta}^{2}s_{\alpha}^{2}\dd\phi_{2}+s_{\alpha}^{2}s_{\theta}^{2}\dd\phi_{3}\,,\\
e_{2}+\ii e_{5} & =\ee^{3\ii\psi/2}\dd\alpha-\ii \ee^{3\ii\psi/2}c_{\alpha}s_{\alpha}\dd\phi_{1}+\ii \ee^{3\ii\psi/2}c_{\alpha}c_{\theta}^{2}s_{\alpha}\dd\phi_{2}+\ii \ee^{3\ii\psi/2} c_{\alpha}s_{\alpha}s_{\theta}^{2}\dd\phi_{3}\,,\\
e_{4}+\ii e_{3} & =\ee^{3\ii\psi/2}s_{\alpha}\dd\theta-\ii \ee^{3\ii\psi/2}c_{\theta}s_{\alpha}s_{\theta}\dd\phi_{2}+\ii \ee^{3\ii\psi/2}c_{\theta}s_{\alpha}s_{\theta}\dd\phi_{3}\,,
\end{split}
\end{equation}
where $3\psi=\phi_{1}+\phi_{2}+\phi_{3}$. The complex, symplectic and contact structures are defined in terms of the frame in (\ref{frame}). One can check they satisfy the correct algebraic and differential relations (\ref{eq:asigma})--(\ref{scalingomega}).

The marginal deformations are given in terms of a function $f$ which is of charge three under the Reeb vector and the restriction of a holomorphic function on $\bbC^3$. In our parametrisation the Reeb vector field is
\beq
\xi=3\del_\psi = \del_{\phi_1} + \del_{\phi_2} + \del_{\phi_3} \,, 
\eeq
and the coordinates $z_i$ have charge +1 
\beq
{\cal L}_\xi z_i=\ii z_i \,.
\eeq
Thus, $f$ must be a cubic function of the $z_{i}$. An arbitrary cubic holomorphic function on $\bbC^3$ has ten complex degrees of freedom and can be written as
\beq
f = f^{ijk} z_i  z_j  z_k \,,
\eeq
where $f^{ijk}$ is a complex symmetric tensor of $\SU{3}$ with ten complex degrees of freedom. This is the same structure as the superpotential deformation~\eqref{WN4def}. As mentioned before, not all components of $f$ correspond to exactly marginal deformations because we still need to take into account the broken $\SU{3}$ global symmetry. This imposes the further constraint
\begin{equation}
f^{ikl}\bar{f}_{jkl}-\tfrac{1}{3}\delta^{i}_{\phantom{i}j}f^{klm}\bar{f}_{klm}=0 \,,
\end{equation}
which removes eight real degrees of freedom. We can also redefine the couplings using the $\SU{3}$ symmetry to remove another eight real degrees of freedom, leaving a two-complex dimensional space of exactly marginal deformations. Thus, there are two independent solutions
\beq
f_{\beta} \propto z_1 z_2 z_3 \,,
\eeq
and 
\beq
f_{\lambda} \propto z_1^3+z_2^3+z_3^3 \,,
\eeq
corresponding to the $\beta$-deformation and the cubic deformation in \eqref{WN4defbis}. 

The supergravity dual of the $\beta$-deformation was worked out in \cite{LM05}. One can check that using our frame for $\text{S}^{5}$ and taking
\begin{equation}
f_\beta=-\tfrac{3}{2} \gamma z_1 z_2 z_3 \,,
\end{equation}
where $\gamma$ is real, our expression (\ref{G3}) for the three-form fluxes reproduces those in the first-order $\beta$-deformed solution~\cite{LM05}. To generate the complex deformation of LM, we promote $\gamma$ to $\gamma - \ii \sigma$, where both $\gamma$ and $\sigma$ are real. This reproduces the LM fluxes with $\tau = \ii$. The full complex deformation with general $\tau$ can be obtained using the $\SL{2;\bbR}$ frame from \cite{LSW14}.

Unlike the $\beta$-deformation, the supergravity dual of the cubic deformation is known only perturbatively. Aharony et.~al have given an expression for the three-form flux for both the $\beta$ and cubic deformations to first order~\cite{GP01}. Again, one can check that our expression reproduces this flux for both $f_\beta$ and $f_\lambda$.

We saw that the marginal deformations \eqref{degrees} also allow for closed primitive $(1,1)$-forms that do not contribute to the flux. If such terms are not exact -- if they are non-trivial in cohomology -- they give additional marginal deformations. On $\mathbb{CP}^2$, the base of $\text{S}^{5}$, there are no closed primitive $(1,1)$-forms that are not exact, and so the marginal deformations are completely determined by the function $f$.

\subsection{Klebanov--Witten theory}

A similar analysis can be performed for deformations of the Klebanov--Witten theory. In this case the dual geometry is $\text{T}^{1,1}$, the coset space $\SU{2}\times\SU{2}/\Uni{1}$ with the topology of $\text{S}^{2}\times\text{S}^{3}$. T$^{1,1}$ can also be viewed as a $\Uni{1}$ fibration over $\mathbb{CP}^{1}\times\mathbb{CP}^{1}$ with metric~\cite{CO90}
\begin{equation} \label{metricT11}
\dd s^{2}(\text{T}^{1,1})=\tfrac{1}{9}(\dd\psi+\cos\theta_{1}\dd\phi_{1}+\cos\theta_{2}\dd\phi_{2})^{2}+\tfrac{1}{6}\sum_{i=1,2}(\dd\theta_{i}^{2}+\sin^{2}\theta_{i}\dd\phi_{i}^{2}) \,.
\end{equation}
Each $\SU{2}$ acts on one $\mathbb{CP}^{1}$, and the $\Uni{1}$ acts as shifts of $\psi$. The Reeb vector field is
\beq
\xi=3 \del_{\psi} \, .
\eeq
 
As with $\text{S}^5$, a holomorphic function on the cone over $\text{T}^{1,1}$ determines the marginal deformations. In this case, the cone is the conifold, defined by
\begin{equation} \label{conifold}
z_{1}^{2}+z_{2}^{2}+z_{3}^{2}+z_{4}^{2}=0 \,,\qquad z_{i}\in\mathbb{C}^{4} \, .
\end{equation}
The conifold equation can also be written as 
\begin{equation}
\det Z_{ij}=0 \,,
\end{equation}
where $Z_{ij}=\sigma_{ij}^{a}z_{a}$, $\sigma^{a}=(\boldsymbol{\sigma},\ii\id)$ and $\boldsymbol{\sigma}$ are the Pauli matrices. We can choose complex coordinates $A_{\alpha}$ and $B_{\dot\alpha}$ ($\alpha=1,2$), corresponding to each $\mathbb{CP}^1$, which are dual to the chiral fields of the gauge theory
\begin{equation}
Z=\begin{pmatrix}z_{3}+\ii z_{4} & z_{1}-\ii z_{2}\\
z_{1}+\ii z_{2} & -z_{3}+\ii z_{4}
\end{pmatrix}=\begin{pmatrix}A_{1}B_{1} & A_{1}B_{2}\\
A_{2}B_{1} & A_{2}B_{2}
\end{pmatrix} \,.
\end{equation}
The complex coordinates $z_a$ can be parametrised by 
\begin{equation}
\begin{split}
z_{1} & =\frac{1}{2}\biggl(\sin\frac{\theta_{1}}{2}\sin\frac{\theta_{2}}{2}\ee^{\frac{\ii}{2}(\psi-\phi_{1}-\phi_{2})}-\ii\cos\frac{\theta_{1}}{2}\cos\frac{\theta_{2}}{2}\ee^{\frac{\ii}{2}(\psi+\phi_{1}+\phi_{2})}\biggr) \,,\\
z_{2} & =\frac{1}{2\ii}\biggl(\sin\frac{\theta_{1}}{2}\sin\frac{\theta_{2}}{2}\ee^{\frac{\ii}{2}(\psi-\phi_{1}-\phi_{2})}+\ii\cos\frac{\theta_{1}}{2}\cos\frac{\theta_{2}}{2}\ee^{\frac{\ii}{2}(\psi+\phi_{1}+\phi_{2})}\biggr) \,,\\
z_{3} & =\frac{1}{2}\biggl(\cos\frac{\theta_{1}}{2}\sin\frac{\theta_{2}}{2}\ee^{\frac{\ii}{2}(\psi+\phi_{1}-\phi_{2})}+\ii\sin\frac{\theta_{1}}{2}\cos\frac{\theta_{2}}{2}\ee^{\frac{\ii}{2}(\psi-\phi_{1}+\phi_{2})}\biggr) \,,\\
z_{4} & =-\frac{1}{2\ii}\biggl(\cos\frac{\theta_{1}}{2}\sin\frac{\theta_{2}}{2}\ee^{\frac{\ii}{2}(\psi+\phi_{1}-\phi_{2})}-\ii\sin\frac{\theta_{1}}{2}\cos\frac{\theta_{2}}{2}\ee^{\frac{\ii}{2}(\psi-\phi_{1}+\phi_{2})}\biggr) \,,
\end{split}
\end{equation}
from which we see they have charge 3/2 under the Reeb vector field
\begin{equation}
\mathcal{L}_{\xi}z_{a}=\tfrac{3}{2}\ii z_{a} \,.
\end{equation}

We can take the following frame for $\text{T}^{1,1}$
\begin{equation}\label{frameT11}
\begin{split}e_{1} & =\tfrac{1}{3}(\dd\psi+\cos\theta_{1}\dd\phi_{1}+\cos\theta_{2}\dd\phi_{2})\,,\\
e_{2}+\ii e_{5} & =\tfrac{\ii}{\sqrt{6}}\ee^{\ii\psi/2}\dd\theta_{1}+\tfrac{1}{\sqrt{6}}\ee^{\ii\psi/2}\sin\theta_{1}\dd\phi_{1}\,,\\
e_{4}+\ii e_{3} & =\tfrac{\ii}{\sqrt{6}}\ee^{\ii\psi/2}\dd\theta_{2}+\tfrac{1}{\sqrt{6}}\ee^{\ii\psi/2}\sin\theta_{2}\dd\phi_{2}\,.\\
\end{split}
\end{equation}
The complex, symplectic and contact structures are defined in terms of the frame in (\ref{frame}). One can check they satisfy the correct algebraic and differential relations (\ref{eq:asigma})--(\ref{scalingomega}).

The function $f$ defining the marginal deformations is of weight three under the Reeb vector and a restriction of a holomorphic function on the conifold. Thus $f$ must be a quadratic function of the $z_a$, namely
\begin{equation}
f=f^{ab}z_{a}z_{b}=f^{\alpha\beta,\dot{\alpha}\dot{\beta}} A_{\alpha}B_{\dot{\alpha}}A_{\beta}B_{\dot{\beta}} \,,
\end{equation}
where $f^{ab}$ is symmetric and traceless (by condition (\ref{conifold})), or analogously $f^{\alpha\beta,\dot{\alpha}\dot{\beta}}$ is symmetric in $\alpha\beta$ and $\dot{\alpha}\dot{\beta}$. These deformations are the $\SU{2} \times \SU{2}$-breaking deformations in \eqref{WKWdef} and generically give nine complex parameters. We remove six real degrees of freedom when solving the moment maps to account for the broken $\SU{2} \times \SU{2}$ symmetry. The moment maps are precisely the beta function conditions given in \eqref{betaKW}. We can also redefine the couplings using $\SU{2}\times\SU{2}$ rotations to remove another six real degrees of freedom, leaving a three-complex dimensional space of exactly marginal deformations labelled $f_\beta, f_2$ and $f_3$ in (\ref{WemKW}). We have
\beq
\begin{split}
f_{\beta} & \propto z_1^2+z_2^2-z_{3}^{2}-z_{4}^{2} \,, \\
f_{2}  & \propto z_{3}^{2}-z_{4}^{2} \,, \\
f_{3}  & \propto z_{1}^{2}-z_{2}^{2} \,.
\end{split}
\eeq
The first of these is the $\beta$-deformation for the KW theory. 
The supergravity dual of the $\beta$-deformation was worked out in \cite{LM05}. One can check that using our frame for $\text{T}^{1,1}$ and taking 
\begin{equation}
f=\tfrac{1}{3} \ii \gamma (z_{1}^{2}+z_{2}^{2}-z_3^2-z_4^2) \,,
\end{equation}
our expression (\ref{G3}) reproduces the three-form fluxes that appear in the first-order $\beta$-deformed solution~\cite{LM05}. To our knowledge, the fluxes for the other deformations were not known before.

Unlike $\mathbb{CP}^{2}$, $\mathbb{CP}^{1} \times \mathbb{CP}^{1}$ admits a primitive, closed $(1,1)$-form $\delta$ that is not exact (specifically the difference of the K\"ahler forms on each $\mathbb{CP}^1$), giving one more exactly marginal deformation, corresponding to a shift of the $B$-field on the $\text{S}^2$. On the gauge theory side, this corresponds to the $\SU{2} \times \SU{2}$-invariant shift in the difference of the gauge couplings in \eqref{WKWdef}. Together with $h$, coming from the superpotential itself, one finds a five-dimensional conformal manifold.


\subsection{\texorpdfstring{Y$^{p,q}$}{Y(p,q)} gauge theories}

We can repeat the analysis of the Klebanov--Witten theory for the $\mathcal{N}=1$ quiver gauge theories of section \ref{sec:GTYpq}. The dual geometries are the family of Sasaki--Einstein spaces known as $\text{Y}^{p,q}$, which have topology $\text{S}^2\times \text{S}^3$ (recall $0\le q \le p$ and $\text{Y}^{1,0}=\text{T}^{1,1}$). The metric is~\cite{GMSW04b}
\begin{equation}
\begin{split}\label{metricYpq}
\dd s^{2}(\text{Y}^{p,q})&=\tfrac{1}{6}(1-y)(\dd\theta^2+\sin^2 \theta \dd \phi) + w(y)^{-1} q(y)^{-1} \dd y^2 + \tfrac{1}{36} w(y) q(y)(\dd\beta+\cos\theta\dd\phi)^2 \\
&\eqspace+\tfrac{1}{9}\bigl(\dd\psi - \cos\theta \dd \phi + y(\dd\beta + \cos\theta\dd\phi)\bigr)^2 \,,
\end{split}
\end{equation}
where the functions $w(y)$ and $q(y)$ are
\begin{equation}
w(y)=\frac{2(a-y^2)}{1-y} \, , \qqq q(y) = \frac{a-3y^2+2y^3}{a-y^2} \, ,
\end{equation}
and $a$ is related to $p$ and $q$ by
\beq
a=\frac12 - \frac{p^2-3 q^2}{4p^3}\sqrt{4p^2-3 q^2} \ . 
\eeq
The Reeb vector field is
\beq
\xi=3 \del_{\psi} \, .
\eeq
 
As with $\text{S}^5$, a holomorphic function on the cone over $\text{Y}^{p,q}$ determines the marginal deformations. The complex coordinates that define the cone for a generic $\text{Y}^{p,q}$ are known but rather complicated~\cite{BHOP05}. However, we need only the coordinates that can contribute to a holomorphic function with charge $+3$ under the Reeb vector -- fortunately there are only three such coordinates 
\begin{equation}
\begin{split}
b_1 &= \ee^{\ii(\psi-\phi)}\cos^2\frac{\theta}{2}\prod^3_{i=1}(y-y_i)^{1/2} \, , \\
b_2 &= \ee^{\ii(\psi+\phi)}\sin^2\frac{\theta}{2}\prod^3_{i=1}(y-y_i)^{1/2} \, , \\
b_3 &= \ee^{\ii\psi}\sin\frac{\theta}{2}\cos\frac{\theta}{2}\prod^3_{i=1}(y-y_i)^{1/2} \,  .
\end{split}
\end{equation}
The $y_i$ are the roots of a certain cubic equation and are given in terms of $p$ and $q$ as
\begin{equation}
\begin{split}
y_1 &= \tfrac{1}{4}p^{-1}\bigl(2p-3q-(4p^2-3q^2)^{1/2}\bigr) \, , \\
y_2 &= \tfrac{1}{4}p^{-1}\bigl(2p+3q-(4p^2-3q^2)^{1/2}\bigr) \, , \\
y_3 &= \tfrac{3}{2}-y_1-y_2 \, .
\end{split}
\end{equation}
The coordinates $b_a$ actually have charge $+3$ under the Reeb vector
\begin{equation}
\mathcal{L}_{\xi}b_a=3\ii b_a \, ,
\end{equation}
and so the holomorphic function that encodes the marginal deformations will be a linear function of the $b_a$.

We can take the following frame for any $\text{Y}^{p,q}$
\begin{equation}\label{frameYpq}
\begin{split}
e_{1} & =\tfrac{1}{3}\bigl(\dd\psi - \cos\theta\dd\phi + y(\dd\beta+\cos\theta\dd\phi)\bigr)\,,\\
e_{2}+\ii e_{5} & = \ee^{\ii \psi / 2} \left(\frac{1-y}{6}\right)^{1/2}(\dd\theta+\ii\sin\theta\dd\phi) \, , \\
e_{4}+\ii e_{3} & = \ee^{\ii\psi/2} w(y)^{-1/2} q(y)^{-1/2} \bigl(\dd y + \tfrac{1}{6}\ii w(y) q(y) (\dd\beta+\cos\theta\dd\phi)\bigr) \, .
\end{split}
\end{equation}
The complex, symplectic and contact structures are defined in terms of the frame in (\ref{frame}). One can check they satisfy the correct algebraic and differential relations (\ref{eq:asigma})--(\ref{scalingomega}).

The function $f$ defining the marginal deformations is of weight three under the Reeb vector and a restriction of a holomorphic function on the cone. Thus $f$ must be a linear combination of the $b_a$, namely
\begin{equation}
f=f^a b_a \,.
\end{equation}
These deformations are the $\SU{2}$-breaking deformations in \eqref{WYpq} and generically give three complex parameters. We remove two real degrees of freedom when solving the moment maps to account for the broken $\SU{2}$ symmetry (leaving a $\Uni{1}$ unbroken). The moment maps are precisely the beta function conditions given in \eqref{betaYpq}. We can also redefine the couplings using $\SU{2}$ rotations to remove another two real degrees of freedom, leaving a one-complex -dimensional space of exactly marginal deformations. The single independent solution is
\beq \label{fbetaYpq}
f_{\beta} \propto b_3 \, .
\eeq
This is the $\beta$-deformation for the quiver gauge theory. The supergravity dual of the $\beta$-deformation for $\text{Y}^{p,q}$ was worked out in \cite{LM05}. One can check that using the frame for $\text{Y}^{p,q}$ given in \eqref{frameYpq} and taking (\ref{fbetaYpq}), our expression (\ref{G3}) reproduces the three-form fluxes that appear in the first-order $\beta$-deformed solution~\cite{LM05}.  Together with $h$ and $\tau$ (dual respectively to the axion-dilaton and the $B$-field on the $\text{S}^2$), one finds a three-dimensional conformal manifold.


\section{Discussion}

In this paper we have used exceptional generalised geometry to analyse exactly marginal deformations of $\mathcal{N}=1$ SCFTs that are dual to AdS$_5$ backgrounds in type II or eleven-dimensional supergravity. In the gauge theory, marginal deformations are determined by imposing F-term conditions on operators of conformal dimension three and then quotienting by the complexified global symmetry group. We have shown that the supergravity analysis gives a geometric interpretation of the gauge theory results. The marginal deformations are obtained as solutions of moment maps for the generalised diffeomorphism group that have the correct charge under the Reeb vector, which generates the $\Uni1_\text{R}$ symmetry. If this is the only symmetry of the background, all marginal deformations are exactly marginal. If the background possesses extra isometries, there are obstructions that come from fixed points of the moment maps. The exactly marginal deformations are then given by a further quotient by these extra isometries. 

For the specific case of Sasaki--Einstein backgrounds in type IIB we showed how supersymmetric deformations can be understood as deformations of generalised structures which give rise to three-form flux perturbations at first order. Using explicit examples, we showed that our expression for the three-form flux matches those in the literature and the obstruction conditions match the one-loop beta functions of the dual SCFT.

Our analysis holds for any $\mathcal{N}=2$ AdS$_5$ background. It would be interesting to apply it to one of the few examples of non-Sasaki--Einstein backgrounds, such as the Pilch--Warner solution~\cite{PW00b}. This is dual to a superconformal fixed point of $\mathcal{N}=4$ super Yang--Mills deformed by a mass for one of the chiral superfields. Another natural direction would be to apply our analysis to backgrounds dual to SCFTs in other dimensions. For example, one can study $\AdS_4$ backgrounds in M-theory, such as $\AdS_4 \times \text{S}^7$, where the solution-generating technique of Lunin and Maldacena to find the $\beta$-deformation also applies. 

It would be interesting to see whether our approach can be used to go beyond the linearised analysis and find the all-order supergravity backgrounds dual to the deformations; so far only the dual of the $\beta$-deformation has been obtained. With these in hand, one would be able to perform many non-trivial checks of the AdS/CFT correspondence, including calculating the metric on the conformal manifold.

Our formalism has applications other than AdS/CFT. Supersymmetric deformations of the geometry give rise to moduli fields in the low-energy effective action obtained after compactifying on the internal manifold. Determining the number and nature of moduli fields that arise in flux compactifications is difficult in general as we lose many of the mathematical tools used in Calabi--Yau compactifications. In our formalism, fluxes and geometry are both encoded by the generalised structure whose deformations will give all the moduli of the low-energy theory. The generalised geometry points to a new set of tools to understand these deformations, such as generalisations of cohomology and special holonomy.

We hope to make progress on these points in the near future.


\acknowledgments

We would like to thank Amihay Hanany, Alessandro Tomasiello and Alberto Zaffaroni for useful discussions. This work was supported in part by the joint Network Grant DFG/LU/419/9-1 and  EPSRC EP/I02784X/1, the ERC Starting Grant 259133 -- ObservableString, the EPSRC Programme Grant ``New Geometric Structures from String Theory'' EP/K034456/1, the EPSRC standard grant EP/N007158/1, the STFC Consolidated Grant ST/L00044X/1, the Swiss National Science Foundation under project P300P2-158440, a public grant as part of the Investissement d'avenir project, reference ANR-11-LABX-0056-LMH, LabEx LMH and COST Action MP1210 ``The String Theory Universe''. M.~Petrini, D.~Waldram and M.~Gra\~na would also like thank the Mainz Institute for Theoretical Physics, the Galileo Galilei Institute for Theoretical Physics and INFN and the Simons Center for Geometry and Physics for their hospitality and partial support during the completion of this work.

\appendix


\section{\texorpdfstring{$\Ex{6(6)}$}{E6(6)} for type IIB}

In this section we provide details of the construction of $\Ex{6(6)}\times\mathbb{R}^{+}$  generalised geometry for  type IIB supergravity compactified on a five-dimensional  manifold $M$. (For more details and the corresponding construction in eleven-dimensional supergravity see~\cite{CSW11,AW15}.) We decompose the relevant $\Ex{6(6)}$ representations according to a $\GL{5;\bbR}\times\SL{2;\bbR}$ subgroup, where $\SL{2;\bbR}$ is the S-duality group and $\GL{5;\bbR}$ acts on $M$. 

The generalised tangent bundle is
\begin{equation}\label{EGT}
\begin{split}E & \simeq TM\oplus T^{*}M\oplus(T^{*}M\oplus\ext^{3}T^{*}M\oplus\ext^{5}T^{*}M)\oplus\ext^{5}T^{*}M\\
 & \simeq TM\oplus(T^{*}M\otimes S)\oplus\ext^{3}T^{*}M\oplus(\ext^{5}T^{*}M\otimes S) \,,
\end{split}
\end{equation}
where $S$ transforms as a doublet of $\SL{2;\bbR}$. We write sections of this bundle as
\begin{equation}
V=v+\lambda^{i}+\rho+\sigma^{i} \,,
\end{equation}
where $v\in\Gamma(TM)$, $\lambda^{i}\in\Gamma(T^{*}M\otimes S)$, $\rho\in\Gamma(\ext^{3}T^{*}M)$ and $\sigma^{i}\in\Gamma(\ext^{5}T^{*}M\otimes S)$. The adjoint bundle is
\begin{equation}
\begin{split}\ad\tilde{F} & =\mathbb{R}\oplus(TM\otimes T^{*}M)\oplus(S\otimes S^{*})\oplus(S\otimes\ext^{2}TM)\oplus(S\otimes\ext^{2}T^{*}M)\\
 & \eqspace\oplus\ext^{4}TM\oplus\ext^{4}T^{*}M \,.
\end{split}
\end{equation}
We write sections of the adjoint bundle as
\begin{equation}
\cA=l+r+a+\beta^{i}+B^{i}+\gamma+C \,,
\end{equation}
where $l\in\Gamma(\mathbb{R})$, $r\in\Gamma(\End TM)$, etc.  The $\ex{6(6)}$ subalgebra is generated by setting $l=r_{\phantom{a}a}^{a}/3$.

We take $\{\hat{e}_{a}\}$ to be a basis for $TM$ with a dual basis $\{e^{a}\}$ on $T^{*}M$ so there is a natural $\gl{5}$ action on tensors. 
For example, the actions on a vector and a three-form are
\begin{equation}
(r\cdot v)^{a}=r_{\phantom{a}b}^{a}v^{b}\,,\qquad(r\cdot\lambda)_{abc}=-r_{\phantom{d}a}^{d}\lambda_{dbc}-r_{\phantom{d}b}^{d}\lambda_{adc}-r_{\phantom{d}c}^{d}\lambda_{abd}\,.
\end{equation}
Our notation follows \cite{CSW14}. Wedge products and contractions are given by
\begin{equation} \label{conventionswedge}
\begin{split}(v\wedge u)^{a_{1}\dots a_{p+p^{\prime}}} & \coloneqq\frac{(p+p^{\prime})!}{p!p^{\prime}!}v^{[a_{1}\dots}u^{a_{p+1}\dots a_{p+p^{\prime}}]}\,,\\
(\lambda\wedge\rho)_{a_{1}\dots a_{q+q^{\prime}}} & \coloneqq\frac{(q+q^{\prime})!}{q!q^{\prime}!}\lambda_{[a_{1}\dots a_{q}}\rho_{a_{q+1}\dots a_{q+q^{\prime}}]}\,,\\
(v\lrcorner\lambda)_{a_{1}\dots a_{q-p}} & \coloneqq\frac{1}{p!}v^{b_{1}\dots b_{p}}\lambda_{b_{1}\dots b_{p}a_{1}\dots a_{q-p}}\quad\text{if }p\leq q\,,\\
(v\lrcorner\lambda)^{a_{1}\dots a_{p-q}} & \coloneqq\frac{1}{q!}v^{a_{1}\dots a_{p-q}b_{1}\dots b_{q}}\lambda_{b_{1}\dots b_{q}}\quad\text{if }p\geq q\,,\\
(jv\lrcorner j\lambda)_{\phantom{a}b}^{a} & \coloneqq\frac{1}{(p-1)!}v^{ac_{1}\dots c_{p-1}}\lambda_{bc_{1}\dots c_{p-1}}\,,\\
(j\lambda\wedge\rho)_{a,a_{1}\ldots a_{d}} & \coloneqq\frac{d!}{(q-1)!(d+1-q)!}\lambda_{a[a_{1}\ldots a_{q-1}}\rho_{a_{q}\ldots a_{d}]}\, .
\end{split}
\end{equation}

We define the adjoint action of $\cA\in\Gamma(\ad\tilde{F})$ on a generalised vector  $V\in \Gamma(E)$  to be $V^{\prime}= \cA\cdot V$, where the components of $V^{\prime}$ are
\begin{equation}
\begin{split}
v^{\prime} & =lv+r\cdot v+\gamma\lrcorner\rho+\epsilon_{ij}\beta^{i}\lrcorner\lambda^{j}\, ,\\
\lambda^{\prime i} & =l\lambda^{i}+r\cdot\lambda^{i}+a_{\phantom{i}j}^{i}\lambda^{j}-\gamma\lrcorner\sigma^{i}+v\lrcorner B^{i}+\beta^{i}\lrcorner\rho \, ,\\
\rho^{\prime} & =l\rho+r\cdot\rho+v\lrcorner C+\epsilon_{ij}\beta^{i}\lrcorner\sigma^{j}+\epsilon_{ij}\lambda^{i}\wedge B^{j} \, ,\\
\sigma^{\prime i} & =l\sigma^{i}+r\cdot\sigma^{i}+a_{\phantom{i}j}^{i}\sigma^{j}-C\wedge\lambda^{i}+\rho\wedge B^{i} \, .
\end{split}
\label{eq:IIB_adjoint}
\end{equation}
We define the adjoint action of $\cA$ on $\cA^{\prime}$ to be $\cA^{\prime\prime}=[\cA, \cA^{\prime}]$, with components
\begin{equation}\label{IIB_algebra}
\begin{split}l^{\prime} & =\tfrac{1}{2}(\gamma\lrcorner C^{\prime}-\gamma^{\prime}\lrcorner C)+\tfrac{1}{4}\epsilon_{kl}(\beta^{k}\lrcorner B^{\prime l}-\beta^{\prime k}\lrcorner B^{l}) \,,\\
r^{\prime\prime} & =(r\cdot r^{\prime}-r^{\prime}\cdot r)+\epsilon_{ij}(j\beta^{i}\lrcorner jB^{\prime j}-j\beta^{\prime i}\lrcorner jB^{j})-\tfrac{1}{4}\id\epsilon_{kl}(\beta^{k}\lrcorner B^{\prime l}-\beta^{\prime k}\lrcorner B^{l})\\
 & \eqspace+(j\gamma\lrcorner jC^{\prime}-j\gamma^{\prime}\lrcorner jC)-\tfrac{1}{2}\id(\gamma\lrcorner C^{\prime}-\gamma^{\prime}\lrcorner C) \,,\\
a_{\phantom{\prime\prime i}j}^{\prime\prime i} & =(a\cdot a^{\prime}-a^{\prime}\cdot a)_{\phantom{i}j}^{i}+\epsilon_{jk}(\beta^{i}\lrcorner B^{\prime k}-\beta^{\prime i}\lrcorner B^{k})-\tfrac{1}{2}\delta_{\phantom{i}j}^{i}\epsilon_{kl}(\beta^{k}\lrcorner B^{\prime l}-\beta^{\prime k}\lrcorner B^{l})\,,\\
\beta^{\prime\prime i} & =(r\cdot\beta^{\prime i}-r^{\prime}\cdot\beta^{i})+(a\cdot\beta^{\prime}-a^{\prime}\cdot\beta)^{i}-(\gamma\lrcorner B^{\prime i}-\gamma^{\prime}\lrcorner B^{i})\,,\\
B^{\prime\prime i} & =(r\cdot B^{\prime i}-r^{\prime}\cdot B^{i})+(a\cdot B^{\prime}-a^{\prime}\cdot B)^{i}+(\beta^{i}\lrcorner C^{\prime}-\beta^{\prime i}\lrcorner C)\,,\\
\gamma^{\prime\prime} & =(r\cdot\gamma^{\prime}-r^{\prime}\cdot\gamma)+\epsilon_{ij}\beta^{i}\wedge\beta^{\prime j}\,,\\
C^{\prime\prime} & =(r\cdot C^{\prime}-r^{\prime}\cdot C)-\epsilon_{ij}B^{i}\wedge B^{\prime j}\,.
\end{split}
\end{equation}
The cubic invariant for $\Ex{6(6)}$ is
\begin{equation}
c(V,V,V)=-\tfrac{1}{2}(\imath_{v}\rho\wedge\rho+\epsilon_{ij}\rho\wedge\lambda^{i}\wedge\lambda^{j}-2\epsilon_{ij}\imath_{v}\lambda^{i}\sigma^{j})\,.\label{eq:IIB_cubic}
\end{equation}
The $\ex{6(6)}$ Killing form or trace in the adjoint is
\begin{equation}
\begin{split}\tr(\cA,\cA^{\prime}) & =\tfrac{1}{2}\Bigl(\tfrac{1}{3}\tr(r)\tr(r^{\prime})+\tr(rr^{\prime})+\tr(aa^{\prime})+\gamma\lrcorner C^{\prime}+\gamma^{\prime}\lrcorner C+\epsilon_{ij}(\beta^{i}\lrcorner B^{\prime j}+\beta^{\prime i}\lrcorner B^{j})\Bigr)\,.
\end{split}
\label{eq:IIB_Killing}
\end{equation}

We now define the generalisation of the Lie derivative. We introduce the dual generalised tangent bundle $E^{*}$ and define a projection 
\begin{equation}
\times_{\text{ad}}\colon E^{*}\otimes E\rightarrow\ad\tilde{F}.
\end{equation}
We embed the usual derivative operator in the one-form component of $E^{*}$ via the map $T^{*}M\rightarrow E^{*}$. In coordinate indices $M$ one defines
\begin{equation}
\partial_{M}=\begin{cases}
\partial_{m} & \text{for }M=m\,,\\
0 & \text{otherwise}\,.
\end{cases}
\end{equation}
Our choice of projection is
\begin{equation}
\partial\times_{\text{ad}}V=\partial\otimes v+\dd\lambda^{i}+\dd\rho\,.
\end{equation}
The generalised Lie derivative is then defined as
\begin{equation}
\Dorft_{V}W=V^{B}\partial_{B}W^{A}-(\partial\times_{\text{ad}}V)_{\phantom{A}B}^{A}W^{B}\,.
\end{equation}
This can be extended to act on tensors using the adjoint action of $\partial\times_{\text{ad}}V\in\Gamma(\ad\tilde{F})$ in the second term. We will need explicit expressions for the generalised Lie derivative of sections of $E$ and $\ad\tilde{F}$.

The generalised Lie derivative acting on a generalised vector is
\begin{equation}
\begin{split}\Dorft_{V}V^{\prime} & =\mathcal{L}_{v}v^{\prime}+(\mathcal{L}_{v}\lambda^{\prime i}-\imath_{v^{\prime}}\dd\lambda^{i})+(\mathcal{L}_{v}\rho^{\prime}-\imath_{v^{\prime}}\dd\rho+\epsilon_{ij}\dd\lambda^{i}\wedge\lambda^{\prime j})\\
 & \eqspace+(\mathcal{L}_{v}\sigma^{\prime i}-\imath_{v^{\prime}}\dd\sigma^{i}+\dd\rho\wedge\lambda^{\prime i}-\dd\lambda^{i}\wedge\rho^{\prime})\,.
\end{split}
\label{eq:IIB_Dorf_vector1}
\end{equation}
The generalised Lie derivative acting on a section of the adjoint bundle is
\begin{equation}\label{eq:IIB_Dorf_adjoint}
\begin{split}\Dorft_{V} \cA & =(\mathcal{L}_{v}l+\tfrac{1}{2}\gamma\lrcorner\dd\rho+\tfrac{1}{4}\epsilon_{kl}\beta^{k}\lrcorner\dd\lambda^{l})\\
 & \eqspace+(\mathcal{L}_{v}r+j\gamma\lrcorner j\dd\rho-\tfrac{1}{2}\id\gamma\lrcorner\dd\rho+\epsilon_{ij}j\beta^{i}\lrcorner j\dd\lambda^{j}-\tfrac{1}{4}\id\epsilon_{kl}\beta^{k}\lrcorner\dd\lambda^{l})\\
 & \eqspace+(\mathcal{L}_{v}a_{\phantom{i}j}^{i}+\epsilon_{jk}\beta^{i}\lrcorner\dd\lambda^{k}-\tfrac{1}{2}\delta_{\phantom{i}j}^{i}\epsilon_{kl}\beta^{k}\lrcorner\dd\lambda^{l})+(\mathcal{L}_{v}\beta^{i}-\gamma\lrcorner\dd\lambda^{i})\\
 & \eqspace+(\mathcal{L}_{v}B^{i}+r\cdot\dd\lambda^{i}+a_{\phantom{i}j}^{i}\dd\lambda^{j}+\beta^{i}\lrcorner\dd\rho)+(\mathcal{L}_{v}\gamma)\\
 & \eqspace+(\mathcal{L}_{v}C+r\cdot\dd\rho+\epsilon_{ij}\dd\lambda^{i}\wedge B^{j})\,.
\end{split}
\end{equation}
The generalised tangent bundle $E$ is patched such that on overlapping neighbourhoods, $U_{i}\cap U_{j}$, a generalised vector $V\in\Gamma(E)$ is patched by
\begin{equation}
V_{(i)}=\ee^{\dd\Lambda_{(ij)}^{i}+\dd\tilde{\Lambda}_{(ij)}}V_{(j)}\,,
\end{equation}
where $\Lambda_{(ij)}^{i}$ and $\tilde{\Lambda}_{(ij)}$ are locally a pair of one-forms and a three-form respectively, and the action of $\ee^{\dd\Lambda_{(ij)}^{i}+\dd\tilde{\Lambda}_{(ij)}}$ is the exponentiated adjoint action. To define the isomorphism \eqref{EGT} between $E$ and $TM\oplus(T^* M\otimes S)\oplus\ldots$ we need a choice of connection
\begin{equation}
V=\ee^{B^{i}+C}\tilde{V} \, ,
\end{equation}
where the ``untwisted'' generalised vector $\tilde{V}$ is a section of $TM\oplus(T^* M\otimes S)\oplus\ldots$, and $B^i$ and $C$ are two- and four-form gauge potentials, with gauge transformations given by
\beq
B_{(i)}^i = B_{(j)}^i + \dd\Lambda_{(ij)}^i \, , \qquad C_{(i)} = C_{(j)} + \dd\tilde\Lambda_{(ij)} + \tfrac{1}{2} \epsilon_{ij}B_{(j)}^i \wedge \dd\Lambda^j_{(ij)} \, . 
\eeq 
We identify the fields $B^i$ with the NS-NS and R-R two-form potentials, and $C$ with the R-R four-form potential 
\beq
B^{1}=B_2 \ , \qquad B^{2}=C_2 \ , \qquad C = C_4 \, .
\eeq
The gauge-invariant field strengths are then
\begin{equation}
F_{3}^{i}=\dd B^{i},\qqq F_{5}=\dd C-\tfrac{1}{2}\epsilon_{ij}F_{3}^{i}\wedge B^{j}\,.
\label{eq:IIB_flux}
\end{equation}
For calculations, it often proves simpler to work with sections of $TM\oplus(T^* M\otimes S)\oplus\ldots$ and include the connection in the definition of the generalised Lie derivative. Following this convention, throughout this paper the generalised tensors we write down will be ``untwisted''. When $B^i$ and $C$ are included, the generalised Lie derivative simplifies in a manner analogous to the $H$-twisted exterior derivative of generalised complex geometry, $\dd_{H}=\dd-H\wedge$. We define the twisted generalised Lie derivative $\Dorf_{V}$ of a generalised tensor $\mu$ by
\begin{equation}\label{IIB_twisted_Dorf}
\Dorf_{V}\mu\coloneqq\ee^{-B^{i}-C}\Dorft_{\ee^{B^{i}+C}V}(\ee^{B^{i}+C}\mu \, \ee^{-B^i-C})\,.
\end{equation}
We find that $\Dorf_{V}$ has the same form as $\Dorft_{V}$ but includes correction terms involving the fluxes. The net effects of this in \eqref{eq:IIB_Dorf_vector1} and \eqref{eq:IIB_Dorf_adjoint} are the substitutions
\begin{equation}\label{eq:IIB_twisted_dorf1}
\dd\lambda^{i}\rightarrow\dd\lambda^{i}-\imath_{v}F_{3}^{i}\,,\qqq\dd\rho\rightarrow\dd\rho-\imath_{v}F_{5}-\epsilon_{ij}\lambda^{i}\wedge F_{3}^{j}\,.
\end{equation}

\section{Supersymmetry conditions and deformations}
\label{sec:technicalities}

In this appendix we give a detailed discussion of the deformations of the Sasaki--Einstein structure and  of the derivation of  the constraints from supersymmetry. We start with a brief description of the generalised structures and then move to their deformations and the conditions that supersymmetry imposes on them.



\subsection{The generalised structures} 
\label{sec:JKapp}

In studying backgrounds with non-trivial fluxes it is often convenient to rewrite the supersymmetry conditions of ten-dimensional supergravity  as equations on differential forms. In this paper we use the reformulation of the supersymmetry variations proposed in~\cite{AW15}, which recasts the supersymmetric background as an integrable exceptional structure in $\Ex{6(6)}\times\mathbb{R}^{+}$ generalised geometry. The structure is defined by the generalised tensors $K$ and $J_\alpha$ introduced in section \ref{sec:JK}. These are globally defined objects that reduce the structure group of the generalised frame bundle so that there is $\cN =2$ supersymmetry in five dimensions. The latter amounts to the existence of a pair of Killing spinors on the internal manifold. Since the internal spinors transform in the $\rep{8}$ representation of the local group $\USp{8}$, a pair of Killing spinors is invariant under a reduced $\USp{6}$ group. 
 
Computing the tensor product of the Killing spinors, one finds the structures $K$ and $J_\alpha$, transforming in the $\rep{27^\prime}$ and $\rep{78}$ of $\Ex{6(6)}$. One can show that $K$ is left invariant by an $\Fx{4(4)}$ subgroup of $\Ex{6(6)}$.  At a point on the internal manifold, $K$ parametrises the coset $\Ex{6(6)}\times {\mathbb R}^+/\Fx{4(4)}$, equivalent to picking an element in the $\rep{27^\prime}$ of $\Ex{6(6)}$ such that $c(K,K,K)>0$. A choice of $K$ for the whole manifold then defines an $\Fx{4(4)}$ structure.

Similarly the triplet $J_\alpha$, at a point on $M$, parametrises the coset $\Ex{6(6)}\times\mathbb{R}^{+}/\SUstar{6}$, equivalent to picking three elements in the $\rep{78}$ of $\Ex{6(6)}$ that form a highest weight $\su{2}$ subalgebra of $\ex{6(6)}$. A choice of $J_\alpha$ for the whole manifold then defines an $\SUstar{6}$ structure. The space of $\SUstar{6}$ structures is the infinite-dimensional space of sections of $J_\alpha$. This space inherits the hyper-K\"ahler structure from the coset at a point.

The normalisations of $K$ and $J_\alpha$ are fixed by the $\Ex{6(6)}$ cubic invariant \eqref{eq:IIB_cubic} and the trace in the adjoint representation \eqref{eq:IIB_Killing}
\begin{equation}
c(K,K,K) = \kappa^2 \,, \qqq \tr(J_\alpha J_\beta) =-\kappa^{2}\delta_{\alpha\beta} \, . 
\end{equation}
The form of the $\Ex{6(6)}$-invariant volume $\kappa^{2}$ depends on the compactification ansatz. For type II compactifications in the string frame of the form
\begin{equation}
g_{10}=\ee^{2\Delta}g_{10-d}+g_{d}\,,
\end{equation}
the invariant volume includes a dilaton dependence and is given by
\begin{equation}
\kappa^{2}=\ee^{-2\phi}\ee^{(8-d)\Delta}\sqrt{g_{d}}\,.\label{eq:IIB_inv_vol}
\end{equation}
For the Sasaki-Einstein backgrounds we consider, this is simply $\kappa^2=\vol_5$.

The two structures are compatible and together define a $\USp{6}$ structure if they satisfy
\beq
J_\alpha\cdot K=0 \,,    
\eeq
where $\cdot$ is the adjoint action \eqref{eq:IIB_adjoint} on a generalised vector.


\subsection{Embedding of the linearised deformations in generalised geometry}

In this section we will justify the choice of \eqref{AAtform} for the linearised deformation. As already mentioned, $K$ is left invariant by an $\Fx{4(4)}$ subgroup of $\Ex{6(6)}$ while the triplet $J_\alpha$ is left invariant by $\SUstar{6}$. Together $J_\alpha$ and $K$ are invariant under a common $\USp{6}$ subgroup. We argued in section \ref{K_left_inv} that the dual of marginal deformations should leave $K$ invariant, but modify the $J_\alpha$. This means that at a point on the internal manifold they must be elements of the coset $\Fx{4(4)}\times\mathbb{R}^{+}/\USp{6}$. The $\rep{52}$ (adjoint) representation of $\Fx{4(4)}$ decomposes under
$\USp{6}\times\SU{2}$ as 
\beq
\rep{52} = (\rep{1},\rep{3})\oplus(\rep{21},\rep{1})\oplus(\rep{14},\rep{2}) \,.
\eeq
The first term corresponds to the triplet $J_\alpha$ and its action simply rotates the $J_\alpha$ among themselves. The second term is the adjoint of $\USp6$, which leaves both $K$ and $J_\alpha$ invariant. Therefore, the deformations are in the $(\rep{14},\rep{2})$ and form a doublet under the $\SU{2}$ defined by $J_\alpha$. We can choose them to be eigenstates of $J_{3}$ 
\begin{equation}
[J_{3},\cA_{\pm \lambda}] = \pm\ii \lambda \kappa \cA_{\pm \lambda} \,.
\end{equation}
The non-trivial eigenstates correspond to  $\lambda = 0, 1, 2$. From the $\SU{2}$ algebra (\ref{su2algJa}) we see that the eigenstates with  $ \lambda = 2$ are $J_\pm$ themselves. The eigenstates with eigenvalue zero are in $\USp{6}$, or in other words they leave $J_\alpha$ and $K$ invariant, and we will therefore not consider them. 
To simplify notation we will call  the $\lambda=\pm1$ eigenstates $\cA_\pm$. We note that we can generate an eigenstate with eigenvalue $-\ii \kappa$ from $\cA_{+}$ by acting with $J_+$, as the Jacobi identity implies
\beq \label{JA}
[J_3,\kappa^{-1}[ J_\pm, \cA_\pm]] = \mp \ii \kappa [ J_\pm, \cA_\pm] \,.
\eeq
We also note that complex conjugation also gives the eigenstate with opposite eigenvalue. Since $\Dorft_K$ commutes with the action of $J_3$ we can also label states by their R-charge as in~\eqref{ARcharge}, so that we have doublets
\begin{equation}
   \cA = \begin{pmatrix} \cA^{(r)}_- \\ \cA^{(r-2)}_+ \end{pmatrix} \, ,
   \qquad r \geq 0 \, . 
\end{equation}
We have chosen $r\geq0$ for definiteness. Those doublets with $r\leq0$ will be related by complex conjugation. (Note this convention leads to a slight over-counting for $0\leq r\leq2$, since the doublets with charge $r$ have complex conjugates with charge $-r+2$. However, it is the most convenient form to adopt for out purposes.) 

To compute the eigenstates with $\lambda=1$ it helps to note that the $\Ex{6(6)}$ action of $J_3$ acts separately on $\{B^i,\beta^i\}$, $a^i_{\phantom{i}j}$ and $\{r,C,\gamma,l\}$ (see \eqref{IIB_algebra}). Using this we can organise the eigenstates as
\begin{align}
\Ac_{+} &=  B^i + \beta^i \, , &\Ac_{-} = [J_+, \Ac_+]&= r +  C +  \gamma +  l \, ,\label{Ap} \\ 
\Ah_{+} &=  a^i_{\phantom{i}j} \, , &\Ah_{-}= [J_+, \Ah_+]&= B'^i + \beta'^i \, .\label{Ap2}
\end{align}
As complex conjugation gives the eigenstate with opposite eigenvalue, using this basis, the modes $\{\Ac_+, \Ac_-^*, \Ah_+, \Ah_-^*\}$ fill out the possible $+\ii\kappa$ eigenstates. In fact we will find that, with this basis, imposing $r\geq0$ actual restricts to only $\Ac_+$ and $\Ah_+$. 

One can use the forms defining the $\SU{2}$ structure on a SE manifold -- $\Omega$, $\omega$ and $\sigma$ -- and the corresponding vectors to decompose the eigenstates. It is straightforward to verify that the eigenstate $\Ac_{+}$ is given by
\beq
\label{Api}
\Ac_{+}   =  - \tfrac12 \ii \bar u^i\bigl[ f \bar{\Omega} +2 (p\omega+ \ot+ \sigma\wedge \bar{\nu}) \bigr]  
-\tfrac12 \bar u^i \bigl[ f \alpha -2 (p \omega^\sharp - \ot^\sharp - \xi \wedge \bar{\upsilon}  ) \bigr] \,,
\eeq
where the vector $u^i$ is defined in \eqref{u},  $\bar{\nu}$ is a (0,1)-form, $\bar{\upsilon}$ is a (1,0)-vector on the base, $\ot$ is a primitive (1,1)-form on the base, and $p$ and $f$ are arbitrary complex functions on the SE manifold. The $\omega^\sharp$ and $\ot^\sharp$ terms in the bi-vector are obtained from the two-forms by raising indices with the metric $g^{mn}$.

The requirement that the deformation leaves $K$ invariant ($\Ac_+ \cdot K = 0$) translates to constraints on the components of $\Ac_+$, namely
\begin{align}
\sigma \wedge \omega \wedge B^i =0 \,, \qqq \imath_{\xi} B^{i} = \beta^{i} \lrc (\sigma \wedge \omega) \, , 
\end{align}
which impose $p =0$ and $ \bar{\upsilon}=\bar{\nu}^{\sharp}$. 
Thus the $\Ac_{+}$ deformation that leaves $K$ invariant is
\beq
\label{Apibis}
\Ac_{+}   =  -\tfrac12 \ii \bar u^i \bigl[ f \bar{\Omega} +2 ( \ot+ \sigma\wedge \bar{\nu}) \bigr]  
-\tfrac12 \bar u^i \bigl[ f \alpha +2 ( \ot+ \xi \wedge \bar{\nu} ) \bigr] \,,
\eeq
where we have omitted the vector symbols $\sharp$ and it is understood that all terms in the bi-vector part are obtained by raising the $\GL{5}$ indices of the corresponding forms with the metric $g^{mn}$. Note that the two-form and bi-vector components are related by
\beq \label{relBbeta}
B^{i} = - \epsilon^i_{\phantom{i}j}  ( g \beta^j g) \, , 
\eeq
where we lower the indices of the bi-vector with the undeformed metric $g$. The $\Ac_-$ mode in the same multiplet as $\Ac_+$ is given by $\Ac_-=\kappa^{-1}[J_+,\Ac_+]$ and has the following form
\begin{equation}\label{Acrminus}
\begin{split}
\Ac_{-} &=\Bigl(2\ii f^\prime \id_4-\ii f^\prime \id+\ii\bigr(j\bar{\alpha}\lrcorner j(\hat{\omega}^\prime+\sigma\wedge\bar{\nu}^\prime)+j(\hat{\omega}^\prime+\xi\wedge\bar{\nu}^\prime)\lrcorner j\Omega\bigr)\Bigr)\\
&\eqspace+(\tfrac{1}{2}f^\prime\Omega\wedge\bar{\Omega}+\Omega\wedge\sigma\wedge\bar{\nu}^\prime)+(\tfrac{1}{2}f^\prime\bar{\alpha}\wedge\alpha+\bar{\alpha}\wedge\xi\wedge\bar{\nu}^\prime)+\ii f^\prime \, ,
\end{split}
\end{equation}
where we should regard $f^\prime$ as distinct from $f$.

Similarly, we can construct the $\Ah_{+}$ deformation that leaves $K$ invariant. It has only $a^i_{\phantom{i}j}$ components, given by
\begin{equation}\label{Ahatdef}
\Ah_+ = -\tfrac{1}{2} \tilde{f} \bar{u}^i \bar{u}_j \,.
\end{equation}
The $\Ah_-$ mode in the same multiplet as $\Ah_+$ is given by $\Ah_{-}=\kappa^{-1}[J_+,\Ah_+]$ and has the following form
\begin{equation}\label{Ahatminus}
\Ah_-   =  (- \tfrac12 \ii \bar u^i \tilde{f}^\prime \Omega) + 
(-\tfrac12 \bar u^i \tilde{f}^\prime \bar{\alpha}) \,,
\end{equation}
where again we should regard $\tilde{f}^\prime$ as distinct from $\tilde{f}$. We see this is of the form $B^i + \beta^i$ as expected from \eqref{Ap2}.


\subsection{Supersymmetry conditions}\label{susy_section}

We are interested in deformations of the Sasaki--Einstein background that preserve supersymmetry. This is equivalent to requiring that the deformed structures are integrable, that is the new $J_\alpha$ and $K$ must satisfy \eqref{mmconditions} and
 \eqref{genLieJa}. At linear order in the deformation these conditions reduce to
 \begin{align}
\label{deltamuaBbis}
\delta \mu_\alpha(V)  = \int \kappa \tr  (J_\alpha, \Dorf_V \cA ) &=0 \qquad \forall \, V \in \rep{27^\prime} \,, \\
\label{genLieJafirstbis}
 [ \Dorf_K \cA, J_\alpha ] &= 0  \, .
\end{align} 
As we want the deformed structures to be real, we take the deformation to be $\mathcal{A}=\re \mathcal{A}_+$, where $\re \mathcal{A}_+ = \frac{1}{2}(\mathcal{A}_+ + \mathcal{A}_+^*)$. In this section we give the derivation of the constraints that these equations impose on the defomations $\Ac_+$. For the other deformations we give only the final results for the constraints, which can be derived in a similar fashion.


\subsubsection{Moment map conditions}
\label{sec:mmtech}

Let us first conside the deformation $\Ac_+$ and the conditions from $\delta \mu_3=0$. Given the form of $J_3$ \eqref{KJSE}, only the $a^{i}_{\phantom{i}j}$, $r^{m}_{\phantom{m}n}$, $C_{mnpq}$ and $\gamma^{mnpq}$ components of the generalised Lie derivative contribute. The relevant terms are
\begin{equation}
\begin{split}
\Dorf_V \Ac_{+} & = ( \epsilon_{ij}j \beta^i  \lrcorner j\dd\lambda^{j}-\tfrac{1}{4} \id \epsilon_{kl} \beta^{k} \lrcorner \dd\lambda^{l} ) \\
& \eqspace +( \epsilon_{jk}\beta^{i}\lrcorner\dd\lambda^{k} - \tfrac{1}{2}\delta_{\phantom{i}j}^{i}\epsilon_{kl}\beta^{k}\lrcorner\dd\lambda^{l}) + ( \epsilon_{ij}\dd\lambda^{i}\wedge B^{i} ) \\
&  = -[\dd\lambda^{i}, \Ac_{+}] \, ,
\end{split}
\end{equation}
where $B^i$ and $\beta^i$ are the two-form and bi-vector components of $\Ac_{+}$. We use this and rearrange the trace to give
\begin{equation}
\label{m3eq}
 \int\kappa\tr( J_{3},\Dorf_V  \Ac_{+} )  
  \propto \int\kappa\tr\bigl( J_{3}, [ \dd\lambda^{i},   \Ac_{+} ] \bigr) \propto \int\kappa\tr\bigl(\dd\lambda^{i},[ J _{3},   \Ac_{+} ] \bigr) \, ,
\end{equation}
with a similar expression for $\Ac_+^*$. Using that $\Ac_{+}$ is an eigenstate of $J_3$ with eigenvalue $+\ii \kappa$ and the form of the trace \eqref{eq:IIB_Killing}, this simplifies to
\begin{equation}
\begin{split}
\int\kappa\tr\bigl(\dd\lambda^{i},[ J _{3},   \Ac_{+} ] \bigr) & \propto \int\kappa^{2}\epsilon_{i j}  \beta^{i}  \lrcorner\dd\lambda^{j}\\
 & \propto \int\epsilon_{i j} \dd( \beta^{i}  \lrcorner \vol_{5}) \wedge \lambda^{j} \, ,
\end{split}
\end{equation}
where we have used $\vol_5 (\beta^i \lrcorner \dd\lambda^j) \propto (\beta^i \lrcorner \vol_5)\wedge\dd\lambda^j$. When combined with the contribution from $\Dorf_V \Ac_{+}^*$, this should hold for arbitrary $\lambda^{j}$ and so we require
\begin{equation}
\dd\Bigl[\bigr(\beta^{i}-(\beta^{i})^*\bigl)  \lrcorner \vol_{5}\Bigr]=0 \,.
\end{equation}
Using the explicit form of $\Ac_{+}$ \eqref{Apibis}, this condition gives
\begin{equation}
\begin{split}
\partial(\bar{\nu}\lrcorner\Omega) & = 0 \, , \\
\partial\ot & = 0 \, , \\
\partial f\wedge\bar{\Omega} + \tfrac{3}{2}\ii\bar{\Omega}\wedge(\bar{\nu}\lrcorner\Omega) & = 2\bar{\partial}\ot+ \tfrac{1}{2}\bar{\Omega}\wedge\mathcal{L}_{\xi}(\bar{\nu}\lrcorner\Omega) \, .
\end{split}
\end{equation}
The analysis of $\delta\mu_{+}$ follows from similar manipulations. For $\delta\mu_{+}$ there are terms that must vanish for arbitrary $v$ and $\rho$. The terms in $\rho$ give
\begin{equation}
\begin{split}
2\partial f  & = \mathcal{L}_{\xi}(\bar{\nu}\lrcorner\Omega) \, , \\
\bar{\partial}f & = 0 \, , \\
\partial(\bar{\nu}\lrcorner\Omega) & = 0 \, , \\
\bar{\partial}(\bar{\nu}\lrcorner\Omega) & = - 4 f \omega \, .
\end{split}
\end{equation}
The terms in $v$ give
\begin{equation}
\begin{split}
\bar{\partial}\bar{\nu} & = - 2 \ii f \bar{\Omega} \, , \\
\bar{\partial}f & = 0 \, , \\
4 \omega \wedge \bar{\nu} + 4 \bar{\partial}\ot+\tfrac{1}{2}\bar{\Omega}\wedge\mathcal{L}_{\xi}(\bar{\nu}\lrcorner\Omega) & = 2 \ii \bar{\Omega}\wedge(\bar{\nu}\lrcorner\Omega) + \bar{\Omega}\wedge\partial f \, .
\end{split}
\end{equation}

Taken together, the moment map conditions on the deformation $ \Ac_{+}$ are
\begin{align}
\label{eq:KR1}\partial(\bar{\nu}\lrcorner\Omega) & = 0 \, , \\
\label{eq:KR2}\partial\ot& = 0 \, , \\
\label{eq:KR3}2\partial f  & = \mathcal{L}_{\xi}(\bar{\nu}\lrcorner\Omega) \, , \\
\label{eq:KR4}\bar{\partial}f & = 0 \, , \\
\label{eq:KR5}\bar{\partial}(\bar{\nu}\lrcorner\Omega) & = - 4 f \omega \, , \\
\label{eq:KR6}\bar{\partial}\bar{\nu} & = - 2 \ii f \bar{\Omega} \, , \\
\label{eq:KR7}\bar{\partial}\ot& = -3\omega\wedge\bar{\nu} \, .
\end{align}
Note that we have simplified some expressions using
\begin{equation}
4\omega\wedge\bar{v} = - \ii \bar{\Omega}\wedge(\bar{v}\lrcorner\Omega) \, , \qqq \omega\wedge(\bar{v}\lrcorner\Omega) = - \ii \Omega\wedge\bar{v} \, ,
\end{equation}
where $\bar{v}$ is an arbitrary (0,1)-form with respect to $I$. 

We want to solve the system \eqref{eq:KR1}--\eqref{eq:KR7} of differential equations to derive the form of the deformation. From \eqref{eq:KR1} we know $\bar{\nu}\lrcorner \Omega$ is closed under $\partial$, and so it may be written as the sum of a $\partial$-closed term and a $\partial$-exact term. However, we also have $\text{H}^{1,0}_{\partial}(M)={0}$ for a five-dimensional Sasaki--Einstein space $M$, and so only a $\partial$-exact term is needed. We make an ansatz
\beq
\label{nuh}
\bar{\nu} \lrcorner\Omega= -\frac{2\ii}{q}\partial f , 
\eeq
where $f$ has a well-defined scaling under $\xi$, $\mathcal{L}_{\xi} f = \ii q f$, and $q$ is non-zero.\footnote{If $q=0$ and $f$ is holomorphic, $f$ is necessarily constant. But from \eqref{eq:KR6}, a constant $f$ requires $\bar\Omega$ to be $\bar\partial$-exact, which is not true. The only solution to the differential conditions for constant $f$ is $f=0$, and so we do not need to consider the case of $q=0$} Next \eqref{eq:KR5} gives
\begin{equation}
\begin{split}
\bar{\partial}(\bar{\nu} \lrcorner\Omega) & =-\frac{2\ii}{q}\bar{\partial}\partial f = 2\ii\partial\bar{\partial} f - 4 f \omega  \equiv -4 f \omega \, .
\end{split}
\end{equation}
We can solve this by taking $f$ to be holomorphic, which also solves \eqref{eq:KR4}. The ansatz for $\bar{\nu}\lrcorner \Omega$, together with the scaling under $\xi$ and holomorphicity of $f$ are enough to satisfy \eqref{eq:KR3}.

We can invert \eqref{nuh} and write $\bar{\nu}$ as 
\beq
\bar{\nu} = \frac{\ii}{2q}\partial f \lrcorner\bar{\Omega} \, . 
\eeq
Then \eqref{eq:KR6} is automatically satisfied
\begin{equation}
\begin{split}
\bar{\partial}\bar{\nu} & =\frac{\ii}{2q}\bar{\partial}(\partial f \lrcorner\bar{\Omega})  =\frac{\ii}{2q}(-4q f \bar{\Omega})  \equiv-2\ii f \bar{\Omega} \, , 
\end{split}
\end{equation}
where we have used $\bar{\partial}(\partial f \lrcorner\bar{\Omega})=-4 q f \bar{\Omega}$ for a holomorphic function $f$.\footnote{In general one has $\partial(\bar{\partial}f\lrcorner\Omega)=\tfrac{1}{2}(q^2+4q-\Delta_0)f\Omega$ and $\bar{\partial}(\partial f\lrcorner\bar{\Omega})=\tfrac{1}{2}(q^2-4q-\Delta_0)f\bar{\Omega}$ for a function satisfying $\Delta f = \Delta_0 f$ and $\mathcal{L}_\xi f = \ii q f$~\cite{EST13}.}

If we take $\ot=\tfrac{1}{4q(q-1)}\partial(\partial f \lrcorner\bar{\Omega})+\delta$, \eqref{eq:KR7} becomes 
\begin{equation}
\begin{split}
\bar{\partial}\ot & =\bar{\partial}\bigl(\tfrac{1}{4q(q-1)}\partial(\partial f \lrcorner\bar{\Omega})+\delta\bigr) \\
 & =\tfrac{1}{4q(q-1)}\bigl(-\partial\bar{\partial}(\partial f \lrcorner\bar{\Omega})-2\omega\wedge\mathcal{L}_{\xi}(\partial f \lrcorner\bar{\Omega})\bigr)+\bar{\partial}\delta \\
 & =\tfrac{1}{q-1}\partial f \wedge\bar{\Omega}-\ii\tfrac{q-3}{2q(q-1)}\omega\wedge(\partial f \lrcorner\bar{\Omega})+\bar{\partial}\delta \\
 & =-\tfrac{3\ii}{2q}\omega\wedge(\partial f \lrcorner\bar{\Omega})+\bar{\partial}\delta \\
 & \equiv-3\omega\wedge\bar{\nu} \, ,
\end{split}
\end{equation}
implying $\bar{\nu} = \frac{\ii}{2q}\partial f \lrcorner\bar{\Omega}$, in agreement with above, and $\bar{\partial}\delta=0$. Finally, \eqref{eq:KR2} implies $\partial\delta=0$.

Taken together, these determine the $\Ac_+$ solutions of the moment map equations. For example, the two-form component of $\Ac_{+}$ is
\begin{equation}
\label{Z1deformB}
B^i  =  - \tfrac12 \ii\bar u^i \left[  f \bar \Omega  + \tfrac{1}{2q(q-1)}\partial(\partial f \lrcorner\bar{\Omega}) +  \tfrac{\ii}{q} \sigma \wedge (\partial f \lrcorner\bar{\Omega})    \right]  - \ii\bar u^i \delta  \,  ,
 \end{equation}
where $f$ is holomorphic with respect to $\partial$ (and hence has charge $q\geq 0$ under the Reeb vector) and $\delta$ is $\partial$- and $\bar{\partial}$-closed (and hence has charge zero). The bi-vector component is determined from this using \eqref{relBbeta}. Notice that $f$-dependent terms and $\delta$ are independent of each other, so we really have two eigenmodes within this expression. In fact, this solution to the moment map equations corresponds to the $\mathcal{A}_{+}^{(r-2)}$ modes with $r\geq0$ labelled by $f$ and $\delta$ in \eqref{defmult}.

Consider now the deformations $\Ah_+$ in \eqref{Ahatdef}. A similar analysis of the moment maps gives
\begin{equation}
\bar\partial \tilde{f}=0 \,,
\end{equation}
so $\tilde{f}$ is holomorphic (and hence has charge $q\geq0$ under the Reeb vector).  This solution corresponds to the $\mathcal{A}_{+}^{(r-2)}$ modes with $r\geq2$ labelled by $\tilde{f}$ in \eqref{defmult}. 

So far we have examined $\Ac_+$ and $\Ah_+$, which correspond to the $\mathcal{A}_{+}^{(r-2)}$ modes in \eqref{defmult} and are parametrised by the holomorphic functions $f$ and $\tilde{f}$, and a $\partial$- and $\bar{\partial}$-closed (1,1)-form $\delta$. Now we comment on the $\mathcal{A}_{-}^{(r)}$ modes, defined by $\mathcal{A}^{(r)}_- =\kappa^{-1} [J_+,\mathcal{A}^{(r-2)}_+]$. Naively, one might think we should solve the moment maps from scratch for an $\mathcal{A}^{(r)}_- $ deformation. For example, the deformation would be calculated using the generic form of $\Ac_-$, given by \eqref{Acrminus}, and would then lead to differential conditions on the components of $\Ac_+$ from which $\Ac_-$ is generated. Fortunately, given a solution $\mathcal{A}_+$ to the deformed moment maps \eqref{deltamuaBbis}, one can show that $\mathcal{A}_- =\kappa^{-1} [J_+,\mathcal{A}_+]$ is automatically a solution too. The components of $\mathcal{A}_-$ are determined by $\mathcal{A}_+$ and the differential conditions on the components of $\mathcal{A}_-$ reduce to the differential conditions on $\mathcal{A}_+$ that we have already given. For example, we have seen that $\Ac_+$ is completely determined by a holomorphic function $f$ and a $\partial$- and $\bar{\partial}$-closed (1,1)-form $\delta$. As $\Ac_- =\kappa^{-1} [J_+,\Ac_+]$ is automatically a solution, it too is determined by a holomorphic function $f^\prime$ and a $\partial$- and $\bar{\partial}$-closed (1,1)-form $\delta^\prime$. Similarly $\Ah_+$ will be determined by holomorphic function $\tilde{f}'$. Here, we should note, however, because of our slight over-counting, the $r=2$ case with constant $f'$ is actually the complex conjugate of the $r=0$ case of $\Ac_+$.


\subsubsection{Lie derivative along \texorpdfstring{$K$}{K}}

At first order in a generic deformation $\cA \in \rep{78}$ of $\Ex{6(6)}$, the generalised Lie derivative condition is given by \eqref{genLieJafirst}. It is straightforward to check that the commutators are non-zero for both $J_+$ and $J_3$, and so the condition reduces to $\Dorf_K \cA=0$. From \eqref{K_Lie}, we know that the generalised Lie derivative along $K$ reduces to the conventional Lie derivative along $\xi$, and so the deformation condition is simply
\begin{equation}\label{marginal}
\mathcal{L}_\xi \cA=0 \, .
\end{equation}
We see that the deformation must have scaling dimension zero under the Reeb vector field. Using the explicit form of $\Ac_+$ and $\Ah_+$, we find $f$ is charge +3 and $\tilde{f}$ is charge zero (which together with $\bar{\partial}{\tilde{f}}=0$ implies $\tilde{f}$ is constant). We also have  $\delta$ is charge zero, which is consistent with it being $\del$- and $\bar{\del}$-closed. This agrees with~\eqref{degrees}. These are precisely the conditions for the deformations to be marginal.


\subsection{Generalised metric}
\label{sec:genmet}

We have deformed the geometry by two-forms and bi-vectors, but the bosonic fields of type II supergravity do not include bi-vectors. As is typical in generalised complex geometry, acting on the bosonic fields, the bi-vector deformation can be traded for deformations by a gauge potential. We first construct the generalised metric and then give the dictionary for translating a bi-vector deformation into a two-form deformation.

A generalised metric defines a $\USp{8}$ structure. $K$ and $J_\alpha$ together define a $\USp{6}$ structure and so also define a generalised metric, though reconstructing the metric from them may be complicated.\footnote{For example, the conventional metric can be recovered from the three- and four-forms defining a $\Gx{2}$ structure, but the relation between the two is not trivial.} For this reason it proves simpler to construct the generalised metric from scratch. For a generalised vector $V$ decomposed as in (\ref{fundGL5}), the generalised metric, in the untwisted basis, is
\begin{equation}\label{gen_metric}
G(V,V)=v^{m}v_{m}+h_{ij}\lambda^{i}_{\phantom{i}m}\lambda^{jm}+\tfrac{1}{3!}\rho_{m_{1}m_{2}m_{3}}\rho^{m_{1}m_{2}m_{3}}+\tfrac{1}{5!}h_{ij}\sigma^{i}_{\phantom{i}m_{1}\ldots m_{5}}\sigma^{j m_{1}\ldots m_{5}} \, ,
\end{equation}
where $h_{ij}$ is the standard metric on $\SL{2}/\SO{2}$ and we have raised/lowered indices using the metric $g_{mn}$.\footnote{We have chosen $C_0=\phi=0$ for the backgrounds we consider, so $h_{ij}$ is simply $\delta_{ij}$.}

The generalised metric defines a $\USp{8}$ structure and so should be left invariant by a $\USp{8}$ subgroup of $\Ex{6(6)}\times\mathbb{R}^{+}$. Using the adjoint action on $V\in\rep{27^{\prime}}$, one can show that $\USp{8}$ is generated by elements of the $\Ex{6(6)}\times\mathbb{R}^{+}$ adjoint satisfying 
\begin{equation}
\begin{aligned}
l& = 0 \, , & a_{ij}& = - a_{ji} \, ,\\
r_{mn} & = -r_{nm} \, , & C_{mnpq} & =-\gamma_{mnpq} \, ,\\
B^{1}_{\phantom{1}mn} & = \beta^{2}_{\phantom{2}mn} \, , & B^{2}_{\phantom{2}mn} & = -\beta^{1}_{\phantom{1}mn} \, .
\end{aligned}
\end{equation}

One can read off the new bosonic background by constructing the deformed generalised metric. The metric, axion-dilaton and four-form R-R potential receive corrections starting at second order. At first order, only the two-form potentials, $B_{2}$ and $C_{2}$, are corrected. If we consider a deformation by a two-form $B^{i}$ and a bi-vector $\beta^{i}$, at first order the resulting two-form deformation is 
\begin{equation} \label{B2C2gen}
B_2=B^{1}-g\beta^{2}g\, , \qqq C_2=B^{2}+g\beta^{1}g \, .
\end{equation}
We see that the bi-vector can be traded for a two-form contribution. This will become more complicated at higher orders in the deformation due to terms from contractions of the bi-vector with the two-form.

As previously mentioned, this procedure is analogous to what is done when trading $\beta$-deformations in generalised complex geometry for metric and $B$-field deformations (see for example equations (3.3) and (3.4) in \cite{ALLP11}).

\subsubsection{Flux induced by deformation}

Using \eqref{B2C2gen} we have that our two-form deformation $\re \Ac_+=B^{i}+\beta^i$ will induce NS-NS and R-R two-form potentials given by 
\begin{equation}
C_{2} = 2 B^{2} \, ,\qqq B_{2} = 2 B^{1} \, .
\end{equation}
The complexified potential is
\begin{equation}
C_{2}-\ii B_{2} = - 2 \ii(B^{1} + \ii B^{2}) \, .
\end{equation}
Using the explicit form of $\Ac_+$ that solves the deformed moment maps \eqref{Z1deformB}, this is
\begin{equation}
C_{2}-\ii B_{2} = - \ii \left[  f \bar \Omega  + \tfrac{1}{2q(q-1)}\partial(\partial f \lrcorner\bar{\Omega}) +  \tfrac{\ii}{q} \sigma \wedge (\partial f \lrcorner\bar{\Omega})  + 2 \delta  \right],
\end{equation}
where $\mathcal{L}_\xi f = \ii q f$.
From \eqref{marginal}, this deformation will correspond to a marginal deformation if $q=3$ and $\delta$ is $\dd$-closed. The complexified potential then simplifies to
\begin{equation}
C_{2}-\ii B_{2} = - \ii \left[  f \bar \Omega  + \tfrac{1}{12}\partial(\partial f \lrcorner\bar{\Omega}) +  \tfrac{\ii}{3} \sigma \wedge (\partial f \lrcorner\bar{\Omega})  + 2 \delta  \right].
\end{equation}
Taking an exterior derivative, the resulting complexified flux $G_{3}=\dd(C_{2}-\ii B_{2})$ is
\begin{equation}
\begin{split}
G_{3} &= -\ii\bigl( \partial f \wedge \bar{\Omega} + \tfrac{1}{12}\bar{\partial}\partial(\partial f\lrcorner\bar{\Omega}) + \ii\tfrac{2}{3} \omega \wedge (\partial f\lrcorner\bar{\Omega}) - \ii\tfrac{1}{3} \sigma \wedge (\partial+\bar{\partial}) (\partial f\lrcorner\bar{\Omega}) \bigr) \\
&= - \tfrac{4}{3}\ii \partial f \wedge \bar{\Omega} + 4 f \sigma \wedge \bar{\Omega} - \tfrac{1}{3} \sigma \wedge \partial (\partial f\lrcorner\bar{\Omega}) \, ,
\end{split}
\end{equation}
where we have used $\dd\delta=0$, $\omega\wedge(\partial f\lrcorner\bar{\Omega})=\ii \partial f \wedge \bar{\Omega}$ and $\bar{\partial} (\partial f\lrcorner\bar{\Omega}) = -12 f \bar{\Omega}$. We stress once more that this flux is valid for marginal deformations of \emph{any} Sasaki--Einstein structure and reproduces the first-order fluxes of the $\beta$-deformation of Lunin and Maldacena \cite{LM05}.


\subsection{Marginal deformations and the axion-dilaton}\label{axiondilaton}

Let us now consider the effect of an $\Ah_+$ deformation. Such a deformation is marginal if $\tilde{f}$ is charge zero under $\xi$, which, when combined with $\bar{\partial}\tilde{f}=0$, implies $\tilde{f}$ is simply a constant complex number. The physical effect of such a marginal deformation can be found from its action on the $\SL{2;\bbR}$ doublets that appear in the generalised metric. For example, the undeformed generalised metric contains terms of the form
\begin{equation}
G(\lambda,\lambda)=\delta_{ij}\lambda^{i}\lrcorner\lambda^{j}+\ldots \, .
\end{equation}
To first order, the deformed generalised metric will then be
\begin{equation}
\begin{split}G(\lambda+\delta\lambda,\lambda+\delta\lambda) & =\delta_{ij}(\lambda^{i}+\delta\lambda^{i})\lrcorner(\lambda^{j}+\delta\lambda^{j})+\ldots\\
 & =(\delta_{ij}+2m_{ij})\lambda^{i}\lrcorner\lambda^{j}+\ldots \, ,
\end{split}
\end{equation}
where
\begin{equation}
m_{ij}=\tfrac{1}{2}\begin{pmatrix}\im \tilde{f} & -\re\tilde{f}\\
- \re\tilde{f} & -\im\tilde{f}
\end{pmatrix},
\end{equation}
which is simply the real part of \eqref{Ahatdef}. We now want to compare this to the form of the generalised metric when the axion-dilaton is included. From \cite{LSW14}, we see this is
\begin{equation}
G(\lambda,\lambda)=h_{ij}\lambda^{i}\lrcorner\lambda^{j}+\ldots,
\end{equation}
where
\begin{equation}
h_{ij}=\ee^{\phi}\begin{pmatrix}C_0^{2}+\ee^{-2\phi} & -C_0\\
-C_0 & 1
\end{pmatrix}.
\end{equation}
Expanding the fields to linear order, we find
\begin{equation}
h_{ij}=\delta_{ij}+\begin{pmatrix}-\phi & -C_0\\
-C_0 & \phi
\end{pmatrix}.
\end{equation}
By comparing this expression with the deformed metric $m_{ij}$, we see we can encode a first-order change in the axion-dilaton by taking $\tilde{f}=C_0 - \ii \phi$.






\end{document}